\documentclass[12pt]{article}

\usepackage[dvips]{graphicx}
\usepackage{amssymb}

\newcommand{\as}  {\mbox{$\alpha_s$}}
\newcommand{\asmz}  {\mbox{$\alpha_s(M_{\mathrm Z})$}}

\newcommand{\be}   {\begin{equation}}
\newcommand{\ee}   {\end  {equation}}
\newcommand{\bea}  {\begin{eqnarray}}
\newcommand{\eea}  {\end  {eqnarray}}
\newcommand{\mbmz} {{\ifmmode m_b(M_{\mathrm Z})\else $m_b(M_{\mathrm Z})$\fi}}
\newcommand{\mbmu} {{\ifmmode m_b(\mu)\else $m_b(\mu)$\fi}}
\newcommand{\gevcc}{{\ifmmode \mathrm{GeV}/c^2\else GeV/$c^2$\fi}}

\begin{document}


\thispagestyle{empty}

\begin{center}
\begin{large}
EUROPEAN ORGANIZATION FOR NUCLEAR RESEARCH (CERN)
\end{large}
\end{center}

\vspace{1cm}
\begin{flushright}
 CERN-EP/2000-093\\
 July 3, 2000
\end{flushright}
\vspace{1cm}

\begin{center}

  {\huge\textbf{A Measurement of the \boldmath$b$\unboldmath-quark Mass\\[3mm]
   from Hadronic Z Decays}} \\[1.5cm]

  {\large The ALEPH Collaboration}\\
\vspace{0.5cm}

\vspace{1.5cm}
\textbf{\large Abstract}
\vspace{0.5cm}
\end{center}

\noindent
Hadronic Z decay data taken with the ALEPH detector at LEP1
are used to measure the three-jet rate as well as moments
of various  event-shape variables.
The ratios of the observables obtained from $b$-tagged events
and from an  inclusive sample are determined.
The mass of the $b$ quark is extracted from a fit 
to the measured ratios using
a next-to-leading order prediction including mass effects.
Taking the first moment of the
$y_3$ distribution, which is the observable with
the smallest hadronization corrections and systematic uncertainties,
the result is
\begin{center}
  $m_b(M_{\mathrm Z})  =  \left[ 3.27 \,\pm\, 0.22 (\mathrm{stat}) \,\pm\, 0.22 (\mathrm{exp}) 
           \,\pm\, 0.38 (\mathrm{had})  \,\pm\, 0.16 (\mathrm{theo})\right] \;\gevcc\quad$.
\end{center}
The measured ratio is alternatively employed to
test the flavour independence of the strong coupling constant
for $b$ and light quarks.  

\vspace{2cm}
\begin{center}
{\textit{submitted to European Physical Journal C}}\\
\vspace{1cm}

\end{center}

\pagestyle{empty}
\newpage
\small
%
\newlength{\saveparskip}
\newlength{\savetextheight}
\newlength{\savetopmargin}
\newlength{\savetextwidth}
\newlength{\saveoddsidemargin}
\newlength{\savetopsep}
\setlength{\saveparskip}{\parskip}
\setlength{\savetextheight}{\textheight}
\setlength{\savetopmargin}{\topmargin}
\setlength{\savetextwidth}{\textwidth}
\setlength{\saveoddsidemargin}{\oddsidemargin}
\setlength{\savetopsep}{\topsep}
%
%
\setlength{\parskip}{0.0cm}
\setlength{\textheight}{25.0cm}
\setlength{\topmargin}{-1.5cm}
\setlength{\textwidth}{16 cm}
\setlength{\oddsidemargin}{-0.0cm}
\setlength{\topsep}{1mm}
\pretolerance=10000
\centerline{\large\bf The ALEPH Collaboration}
\footnotesize
\vspace{0.5cm}
{\raggedbottom
\begin{sloppypar}
\samepage\noindent
R.~Barate,
D.~Decamp,
P.~Ghez,
C.~Goy,
\mbox{J.-P.~Lees},
E.~Merle,
\mbox{M.-N.~Minard},
B.~Pietrzyk
\nopagebreak
\begin{center}
\parbox{15.5cm}{\sl\samepage
Laboratoire de Physique des Particules (LAPP), IN$^{2}$P$^{3}$-CNRS,
F-74019 Annecy-le-Vieux Cedex, France}
\end{center}\end{sloppypar}
\vspace{2mm}
\begin{sloppypar}
\noindent
S.~Bravo,
M.P.~Casado,
M.~Chmeissani,
J.M.~Crespo,
E.~Fernandez,
\mbox{M.~Fernandez-Bosman},
Ll.~Garrido,$^{15}$
E.~Graug\'{e}s,
M.~Martinez,
G.~Merino,
R.~Miquel,
Ll.M.~Mir,
A.~Pacheco,
H.~Ruiz
\nopagebreak
\begin{center}
\parbox{15.5cm}{\sl\samepage
Institut de F\'{i}sica d'Altes Energies, Universitat Aut\`{o}noma
de Barcelona, E-08193 Bellaterra (Barcelona), Spain$^{7}$}
\end{center}\end{sloppypar}
\vspace{2mm}
\begin{sloppypar}
\noindent
A.~Colaleo,
D.~Creanza,
M.~de~Palma,
G.~Iaselli,
G.~Maggi,
M.~Maggi,
S.~Nuzzo,
A.~Ranieri,
G.~Raso,
F.~Ruggieri,
G.~Selvaggi,
L.~Silvestris,
P.~Tempesta,
A.~Tricomi,$^{3}$
G.~Zito
\nopagebreak
\begin{center}
\parbox{15.5cm}{\sl\samepage
Dipartimento di Fisica, INFN Sezione di Bari, I-70126
Bari, Italy}
\end{center}\end{sloppypar}
\vspace{2mm}
\begin{sloppypar}
\noindent
X.~Huang,
J.~Lin,
Q. Ouyang,
T.~Wang,
Y.~Xie,
R.~Xu,
S.~Xue,
J.~Zhang,
L.~Zhang,
W.~Zhao
\nopagebreak
\begin{center}
\parbox{15.5cm}{\sl\samepage
Institute of High Energy Physics, Academia Sinica, Beijing, The People's
Republic of China$^{8}$}
\end{center}\end{sloppypar}
\vspace{2mm}
\begin{sloppypar}
\noindent
D.~Abbaneo,
G.~Boix,$^{6}$
O.~Buchm\"uller,
M.~Cattaneo,
F.~Cerutti,
G.~Dissertori,
H.~Drevermann,
R.W.~Forty,
M.~Frank,
T.C.~Greening,
A.W. Halley,
J.B.~Hansen,
J.~Harvey,
P.~Janot,
B.~Jost,
I.~Lehraus,
P.~Mato,
A.~Minten,
A.~Moutoussi,
F.~Ranjard,
L.~Rolandi,
D.~Schlatter,
M.~Schmitt,$^{20}$
O.~Schneider,$^{2}$
P.~Spagnolo,
W.~Tejessy,
F.~Teubert,
E.~Tournefier,
A.E.~Wright
\nopagebreak
\begin{center}
\parbox{15.5cm}{\sl\samepage
European Laboratory for Particle Physics (CERN), CH-1211 Geneva 23,
Switzerland}
\end{center}\end{sloppypar}
\vspace{2mm}
\begin{sloppypar}
\noindent
Z.~Ajaltouni,
F.~Badaud,
G.~Chazelle,
O.~Deschamps,
A.~Falvard,
P.~Gay,
C.~Guicheney,
P.~Henrard,
J.~Jousset,
B.~Michel,
S.~Monteil,
\mbox{J-C.~Montret},
D.~Pallin,
P.~Perret,
F.~Podlyski
\nopagebreak
\begin{center}
\parbox{15.5cm}{\sl\samepage
Laboratoire de Physique Corpusculaire, Universit\'e Blaise Pascal,
IN$^{2}$P$^{3}$-CNRS, Clermont-Ferrand, F-63177 Aubi\`{e}re, France}
\end{center}\end{sloppypar}
\vspace{2mm}
\begin{sloppypar}
\noindent
J.D.~Hansen,
J.R.~Hansen,
P.H.~Hansen,$^{1}$
B.S.~Nilsson,
A.~W\"a\"an\"anen
\begin{center}
\parbox{15.5cm}{\sl\samepage
Niels Bohr Institute, DK-2100 Copenhagen, Denmark$^{9}$}
\end{center}\end{sloppypar}
\vspace{2mm}
\begin{sloppypar}
\noindent
G.~Daskalakis,
A.~Kyriakis,
C.~Markou,
E.~Simopoulou,
A.~Vayaki
\nopagebreak
\begin{center}
\parbox{15.5cm}{\sl\samepage
Nuclear Research Center Demokritos (NRCD), GR-15310 Attiki, Greece}
\end{center}\end{sloppypar}
\vspace{2mm}
\begin{sloppypar}
\noindent
A.~Blondel,$^{12}$
G.~Bonneaud,
\mbox{J.-C.~Brient},
A.~Roug\'{e},
M.~Rumpf,
M.~Swynghedauw,
M.~Verderi,
\linebreak
H.~Videau
\nopagebreak
\begin{center}
\parbox{15.5cm}{\sl\samepage
Laboratoire de Physique Nucl\'eaire et des Hautes Energies, Ecole
Polytechnique, IN$^{2}$P$^{3}$-CNRS, \mbox{F-91128} Palaiseau Cedex, France}
\end{center}\end{sloppypar}
\vspace{2mm}
\begin{sloppypar}
\noindent
E.~Focardi,
G.~Parrini,
K.~Zachariadou
\nopagebreak
\begin{center}
\parbox{15.5cm}{\sl\samepage
Dipartimento di Fisica, Universit\`a di Firenze, INFN Sezione di Firenze,
I-50125 Firenze, Italy}
\end{center}\end{sloppypar}
\vspace{2mm}
\begin{sloppypar}
\noindent
A.~Antonelli,
M.~Antonelli,
G.~Bencivenni,
G.~Bologna,$^{4}$
F.~Bossi,
P.~Campana,
G.~Capon,
V.~Chiarella,
P.~Laurelli,
G.~Mannocchi,$^{5}$
F.~Murtas,
G.P.~Murtas,
L.~Passalacqua,
\mbox{M.~Pepe-Altarelli}
\nopagebreak
\begin{center}
\parbox{15.5cm}{\sl\samepage
Laboratori Nazionali dell'INFN (LNF-INFN), I-00044 Frascati, Italy}
\end{center}\end{sloppypar}
\vspace{2mm}
\begin{sloppypar}
\noindent
J.G.~Lynch,
P.~Negus,
V.~O'Shea,
C.~Raine,
\mbox{P.~Teixeira-Dias},
A.S.~Thompson
\nopagebreak
\begin{center}
\parbox{15.5cm}{\sl\samepage
Department of Physics and Astronomy, University of Glasgow, Glasgow G12
8QQ,United Kingdom$^{10}$}
\end{center}\end{sloppypar}
\vspace{2mm}
\begin{sloppypar}
\noindent
R.~Cavanaugh,
S.~Dhamotharan,
C.~Geweniger,$^{1}$
P.~Hanke,
G.~Hansper,
V.~Hepp,
E.E.~Kluge,
A.~Putzer,
J.~Sommer,
K.~Tittel,
S.~Werner,$^{19}$
M.~Wunsch$^{19}$
\nopagebreak
\begin{center}
\parbox{15.5cm}{\sl\samepage
Kirchhoff-Institut f\"r Physik, Universit\"at Heidelberg, D-69120
Heidelberg, Germany$^{16}$}
\end{center}\end{sloppypar}
\vspace{2mm}
\begin{sloppypar}
\noindent
R.~Beuselinck,
D.M.~Binnie,
W.~Cameron,
P.J.~Dornan,
M.~Girone,
N.~Marinelli,
J.K.~Sedgbeer,
J.C.~Thompson,$^{14}$
E.~Thomson$^{22}$
\nopagebreak
\begin{center}
\parbox{15.5cm}{\sl\samepage
Department of Physics, Imperial College, London SW7 2BZ,
United Kingdom$^{10}$}
\end{center}\end{sloppypar}
\vspace{2mm}
\begin{sloppypar}
\noindent
V.M.~Ghete,
P.~Girtler,
E.~Kneringer,
D.~Kuhn,
G.~Rudolph
\nopagebreak
\begin{center}
\parbox{15.5cm}{\sl\samepage
Institut f\"ur Experimentalphysik, Universit\"at Innsbruck, A-6020
Innsbruck, Austria$^{18}$}
\end{center}\end{sloppypar}
\vspace{2mm}
\begin{sloppypar}
\noindent
C.K.~Bowdery,
P.G.~Buck,
A.J.~Finch,
F.~Foster,
G.~Hughes,
R.W.L.~Jones,
N.A.~Robertson
\nopagebreak
\begin{center}
\parbox{15.5cm}{\sl\samepage
Department of Physics, University of Lancaster, Lancaster LA1 4YB,
United Kingdom$^{10}$}
\end{center}\end{sloppypar}
\vspace{2mm}
\begin{sloppypar}
\noindent
I.~Giehl,
K.~Jakobs,
K.~Kleinknecht,
G.~Quast,$^{1}$
B.~Renk,
E.~Rohne,
\mbox{H.-G.~Sander},
H.~Wachsmuth,
C.~Zeitnitz
\nopagebreak
\begin{center}
\parbox{15.5cm}{\sl\samepage
Institut f\"ur Physik, Universit\"at Mainz, D-55099 Mainz, Germany$^{16}$}
\end{center}\end{sloppypar}
\vspace{2mm}
\begin{sloppypar}
\noindent
A.~Bonissent,
J.~Carr,
P.~Coyle,
O.~Leroy,
P.~Payre,
D.~Rousseau,
M.~Talby
\nopagebreak
\begin{center}
\parbox{15.5cm}{\sl\samepage
Centre de Physique des Particules, Universit\'e de la M\'editerran\'ee,
IN$^{2}$P$^{3}$-CNRS, F-13288 Marseille, France}
\end{center}\end{sloppypar}
\vspace{2mm}
\begin{sloppypar}
\noindent
M.~Aleppo,
F.~Ragusa
\nopagebreak
\begin{center}
\parbox{15.5cm}{\sl\samepage
Dipartimento di Fisica, Universit\`a di Milano e INFN Sezione di Milano,
I-20133 Milano, Italy}
\end{center}\end{sloppypar}
\vspace{2mm}
\begin{sloppypar}
\noindent
H.~Dietl,
G.~Ganis,
A.~Heister,
K.~H\"uttmann,
G.~L\"utjens,
C.~Mannert,
W.~M\"anner,
\mbox{H.-G.~Moser},
S.~Schael,
R.~Settles,$^{1}$
H.~Stenzel,
W.~Wiedenmann,
G.~Wolf
\nopagebreak
\begin{center}
\parbox{15.5cm}{\sl\samepage
Max-Planck-Institut f\"ur Physik, Werner-Heisenberg-Institut,
D-80805 M\"unchen, Germany\footnotemark[16]}
\end{center}\end{sloppypar}
\vspace{2mm}
\begin{sloppypar}
\noindent
P.~Azzurri,
J.~Boucrot,$^{1}$
O.~Callot,
S.~Chen,
A.~Cordier,
M.~Davier,
L.~Duflot,
\mbox{J.-F.~Grivaz},
Ph.~Heusse,
A.~Jacholkowska,$^{1}$
F.~Le~Diberder,
J.~Lefran\c{c}ois,
\mbox{A.-M.~Lutz},
\mbox{M.-H.~Schune},
\mbox{J.-J.~Veillet},
I.~Videau,$^{1}$
D.~Zerwas
\nopagebreak
\begin{center}
\parbox{15.5cm}{\sl\samepage
Laboratoire de l'Acc\'el\'erateur Lin\'eaire, Universit\'e de Paris-Sud,
IN$^{2}$P$^{3}$-CNRS, F-91898 Orsay Cedex, France}
\end{center}\end{sloppypar}
\vspace{2mm}
\begin{sloppypar}
\noindent
G.~Bagliesi,
T.~Boccali,
G.~Calderini,
V.~Ciulli,
L.~Fo\`{a},
A.~Giassi,
F.~Ligabue,
A.~Messineo,
F.~Palla,$^{1}$
G.~Rizzo,
G.~Sanguinetti,
A.~Sciab\`a,
G.~Sguazzoni,
R.~Tenchini,$^{1}$
A.~Venturi,
P.G.~Verdini
\samepage
\begin{center}
\parbox{15.5cm}{\sl\samepage
Dipartimento di Fisica dell'Universit\`a, INFN Sezione di Pisa,
e Scuola Normale Superiore, I-56010 Pisa, Italy}
\end{center}\end{sloppypar}
\vspace{2mm}
\begin{sloppypar}
\noindent
G.A.~Blair,
G.~Cowan,
M.G.~Green,
T.~Medcalf,
J.A.~Strong,
\mbox{J.H.~von~Wimmersperg-Toeller}
\nopagebreak
\begin{center}
\parbox{15.5cm}{\sl\samepage
Department of Physics, Royal Holloway \& Bedford New College,
University of London, Surrey TW20 OEX, United Kingdom$^{10}$}
\end{center}\end{sloppypar}
\vspace{2mm}
\begin{sloppypar}
\noindent
R.W.~Clifft,
T.R.~Edgecock,
P.R.~Norton,
I.R.~Tomalin
\nopagebreak
\begin{center}
\parbox{15.5cm}{\sl\samepage
Particle Physics Dept., Rutherford Appleton Laboratory,
Chilton, Didcot, Oxon OX11 OQX, United Kingdom$^{10}$}
\end{center}\end{sloppypar}
\vspace{2mm}
\begin{sloppypar}
\noindent
\mbox{B.~Bloch-Devaux},
P.~Colas,
S.~Emery,
W.~Kozanecki,
E.~Lan\c{c}on,
\mbox{M.-C.~Lemaire},
E.~Locci,
P.~Perez,
J.~Rander,
\mbox{J.-F.~Renardy},
A.~Roussarie,
\mbox{J.-P.~Schuller},
J.~Schwindling,
A.~Trabelsi,$^{21}$
B.~Vallage
\nopagebreak
\begin{center}
\parbox{15.5cm}{\sl\samepage
CEA, DAPNIA/Service de Physique des Particules,
CE-Saclay, F-91191 Gif-sur-Yvette Cedex, France$^{17}$}
\end{center}\end{sloppypar}
\vspace{2mm}
\begin{sloppypar}
\noindent
S.N.~Black,
J.H.~Dann,
R.P.~Johnson,
H.Y.~Kim,
N.~Konstantinidis,
A.M.~Litke,
M.A. McNeil,
\linebreak
G.~Taylor
\nopagebreak
\begin{center}
\parbox{15.5cm}{\sl\samepage
Institute for Particle Physics, University of California at
Santa Cruz, Santa Cruz, CA 95064, USA$^{13}$}
\end{center}\end{sloppypar}
\vspace{2mm}
\begin{sloppypar}
\noindent
C.N.~Booth,
S.~Cartwright,
F.~Combley,
M.~Lehto,
L.F.~Thompson
\nopagebreak
\begin{center}
\parbox{15.5cm}{\sl\samepage
Department of Physics, University of Sheffield, Sheffield S3 7RH,
United Kingdom$^{10}$}
\end{center}\end{sloppypar}
\vspace{2mm}
\begin{sloppypar}
\noindent
K.~Affholderbach,
A.~B\"ohrer,
S.~Brandt,
C.~Grupen,$^{1}$
A.~Misiejuk,
G.~Prange,
U.~Sieler
\nopagebreak
\begin{center}
\parbox{15.5cm}{\sl\samepage
Fachbereich Physik, Universit\"at Siegen, D-57068 Siegen,
 Germany$^{16}$}
\end{center}\end{sloppypar}
\vspace{2mm}
\begin{sloppypar}
\noindent
G.~Giannini,
B.~Gobbo
\nopagebreak
\begin{center}
\parbox{15.5cm}{\sl\samepage
Dipartimento di Fisica, Universit\`a di Trieste e INFN Sezione di Trieste,
I-34127 Trieste, Italy}
\end{center}\end{sloppypar}
\vspace{2mm}
\begin{sloppypar}
\noindent
J.~Rothberg,
S.~Wasserbaech
\nopagebreak
\begin{center}
\parbox{15.5cm}{\sl\samepage
Experimental Elementary Particle Physics, University of Washington, Seattle, 
WA 98195 U.S.A.}
\end{center}\end{sloppypar}
\vspace{2mm}
\begin{sloppypar}
\noindent
S.R.~Armstrong,
K.~Cranmer,
P.~Elmer,
D.P.S.~Ferguson,
Y.~Gao,
S.~Gonz\'{a}lez,
O.J.~Hayes,
H.~Hu,
S.~Jin,
J.~Kile,
P.A.~McNamara III,
J.~Nielsen,
W.~Orejudos,
Y.B.~Pan,
Y.~Saadi,
I.J.~Scott,
J.~Walsh,
Sau~Lan~Wu,
X.~Wu,
G.~Zobernig
\nopagebreak
\begin{center}
\parbox{15.5cm}{\sl\samepage
Department of Physics, University of Wisconsin, Madison, WI 53706,
USA$^{11}$}
\end{center}\end{sloppypar}
}
\footnotetext[1]{Also at CERN, 1211 Geneva 23, Switzerland.}
\footnotetext[2]{Now at Universit\'e de Lausanne, 1015 Lausanne, Switzerland.}
\footnotetext[3]{Also at Dipartimento di Fisica di Catania and INFN Sezione di
 Catania, 95129 Catania, Italy.}
\footnotetext[4]{Also Istituto di Fisica Generale, Universit\`{a} di
Torino, 10125 Torino, Italy.}
\footnotetext[5]{Also Istituto di Cosmo-Geofisica del C.N.R., Torino,
Italy.}
\footnotetext[6]{Supported by the Commission of the European Communities,
contract ERBFMBICT982894.}
\footnotetext[7]{Supported by CICYT, Spain.}
\footnotetext[8]{Supported by the National Science Foundation of China.}
\footnotetext[9]{Supported by the Danish Natural Science Research Council.}
\footnotetext[10]{Supported by the UK Particle Physics and Astronomy Research
Council.}
\footnotetext[11]{Supported by the US Department of Energy, grant
DE-FG0295-ER40896.}
\footnotetext[12]{Now at Departement de Physique Corpusculaire, Universit\'e de
Gen\`eve, 1211 Gen\`eve 4, Switzerland.}
\footnotetext[13]{Supported by the US Department of Energy,
grant DE-FG03-92ER40689.}
\footnotetext[14]{Also at Rutherford Appleton Laboratory, Chilton, Didcot, UK.}
\footnotetext[15]{Permanent address: Universitat de Barcelona, 08208 Barcelona,
Spain.}
\footnotetext[16]{Supported by the Bundesministerium f\"ur Bildung,
Wissenschaft, Forschung und Technologie, Germany.}
\footnotetext[17]{Supported by the Direction des Sciences de la
Mati\`ere, C.E.A.}
\footnotetext[18]{Supported by the Austrian Ministry for Science and Transport.}
\footnotetext[19]{Now at SAP AG, 69185 Walldorf, Germany.}
\footnotetext[20]{Now at Harvard University, Cambridge, MA 02138, U.S.A.}
\footnotetext[21]{Now at D\'epartement de Physique, Facult\'e des Sciences de Tunis, 1060 Le Belv\'ed\`ere, Tunisia.}
\footnotetext[22]{Now at Department of Physics, Ohio State University, Columbus, OH 43210-1106, U.S.A.}
%
\setlength{\parskip}{\saveparskip}
\setlength{\textheight}{\savetextheight}
\setlength{\topmargin}{\savetopmargin}
\setlength{\textwidth}{\savetextwidth}
\setlength{\oddsidemargin}{\saveoddsidemargin}
\setlength{\topsep}{\savetopsep}
\normalsize
\newpage
\pagestyle{plain}
\setcounter{page}{1}


\section{Introduction}
 \label{intro}

The advent of next-to-leading order (NLO)
calculations for $\mathrm{e}^+\mathrm{e}^-$ annihilation
into quark pairs, which take full account of quark mass effects 
\cite{rodrigo1, rodrigo2, Bernreuther, Nasonbmass}, has opened up the
possibility for further investigations of QCD, such as measurements of the 
flavour independence of the strong coupling constant 
\cite{Delphibm1, OPALflavindep, SLACflavindep} or measurements
of the running $b$-quark mass at the Z mass scale \cite{Delphibm1, SLACbm1}. 
The $b$-quark mass is one of the fundamental parameters
of the Standard Model Lagrangian, and can in principle be viewed
as a parameter similar to the  coupling constant. The renormalization 
procedure leads to the definition of a running quantity,
the renormalized $b$-quark mass, which is a function of the renormalization
scale. An interpretation as a particle mass is difficult because of  the
fact that quarks are not asymptotically free states. Nevertheless,
a definition of the mass as the pole of the quark propagator is used
frequently, which, depending on the size of long-distance QCD effects,
can be interpreted as the mass of an (almost) free  particle. 

Most of the $b$-quark mass determinations have been performed at
rather low scales \cite{PDG}. It is therefore interesting to 
measure this parameter at a large scale.
Previous measurements have used the ratio of three-jet rates
in $b$- and $uds$-quark decays of the Z boson. For a running $b$-quark
mass of about $3\,\gevcc$ a 3\% deviation from unity is observed 
in this ratio \cite{Delphibm1}. 
This is because the large $b$-quark mass suppresses gluon radiation in a
similar way to the suppression of  bremsstrahlung for muons compared to electrons.

In this analysis a 
set of event-shape observables has been employed  in addition to
the three-jet rate in order to study the $b$-quark mass dependence of the
measured ratios of the observables in $b$ and $uds$ initiated events.
The suppression of gluon radiation by the $b$-quark mass
affects the event-shape distributions, and these 
variables have rather different behaviour with respect to 
the hadronization and next-to-leading order effects. 
The final result is derived using the first moment of the 
$y_3$ distribution, which is the  least affected by
hadronization and next-to-leading order corrections.

In the following section the analysis method is outlined, then a description
of the ALEPH detector is given, followed by a summary of the data
analysis. In Section \ref{data} the results for the measurements
of the ratios of observables in $b$ and $uds$ events are given, which
are then used to determine the $b$-quark mass in Section \ref{massextraction}.
A test of the flavour independence
of the strong coupling constant is described in Section \ref{flavourindep}, 
and the conclusions are given in Section \ref{conclusions}.


\section{The analysis method}
 \label{method}

The running $b$-quark
mass enters into the theoretical prediction for the ratio
\be
 \label{ratio_general}
         R^{\mathrm{pert}}_{bl} = \frac{O_b}{O_l} \quad ,
\ee
where $O_b$ and $O_l$ are infrared and collinear safe
observables at the perturbative level (quarks and gluons) 
for Z decays into $b$ and light ($l=uds$) quark pairs, respectively. The 
measured ratio $R^{\mathrm{meas}}_{bq}$ of the observables in $b$-tagged
events and in an all-flavour inclusive sample
can be related to the
quantities at the parton level via the
following formula:
\be
 \label{correction0}
  R^{\mathrm{meas}}_{bq} = \left(\,
  O_b\, H_b\, D_b\, T_{b}\, {\cal P}_{b}\, +\, 
  O_c\, H_c\, D_c\, T_{c}\, {\cal P}_{c}\, +\,
  O_l\, H_l\, D_l\, T_{l}\, {\cal P}_{l}\,\right)\;/\;
  {O_q^{\mathrm{meas}}}  \quad .
\ee
Here $H_x, D_x, T_x$ are the corrections due to hadronization,
detector effects and tagging, respectively, and 
${\cal P}_{x}$ are the purities of the tagged sample, where
$x$ is the true flavour. The tagging corrections take into
account biases introduced by the flavour tag.
The denominator $O_q^{\mathrm{meas}}$ can be rewritten as
\be
   O_q^{\mathrm{meas}} = R_b\, O_b\, H_b\, D_b +\, 
                    R_c\, O_c\, H_c\, D_c +\, 
                    (1 - R_b - R_c)\, O_l\, H_l\, D_l \quad ,
\ee
where $R_{b(c)}$ is the ratio of the 
partial width of the Z to $b (c)$ quarks and the total hadronic width.
Taking all these ingredients, the relation
between the measured ratio $R^{\mathrm{meas}}_{bq}$ and the ratio of
interest $R^{\mathrm{pert}}_{bl}$ is given by
\be
  \label{correction}
    R^{\mathrm{meas}}_{bq} = \frac{
    R^{\mathrm{pert}}_{bl}\, H_{b/l}\, D_{b/l}\, T_{b}\, {\cal P}_{b}\, +\,
    R^{\mathrm{pert}}_{cl}\, H_{c/l}\, D_{c/l}\, T_{c}\, {\cal P}_{c}\, +\,
                                        T_{l}\, {\cal P}_{l}}{
    R_b\, R^{\mathrm{pert}}_{bl}\, H_{b/l}\, D_{b/l}\,  +\, 
    R_c\, R^{\mathrm{pert}}_{cl}\, H_{c/l}\, D_{c/l}\, + \, (1 - R_b - R_c)}  \quad ,
 \ee
where $H_{b/l} = H_b/H_l$, $D_{b/l} = D_b/D_l$, and the ratio
$R^{\mathrm{pert}}_{cl}$ is set to 1.
A small deviation of this ratio from unity because of the
$c$-quark mass is considered in the systematic studies.
All the correction factors and purities are obtained from
Monte Carlo (MC) simulations. 
Because mainly ratios of corrections are involved,
some systematic uncertainties cancel.

$R^{\mathrm{pert}}_{bl}$ is extracted from the relationship (\ref{correction})
and finally corrected for the contribution of anomalous triangle
diagrams \cite{triangle}, in order to relate $R^{\mathrm{pert}}_{bl}$ to
$R^{\mathrm{pert}}_{bd}$ for which the perturbative calculations have been
performed. The triangle contributions cancel out in the second ratio.
However, they give a contribution of the order of $0.2\%$ to the
first.

The following observables $O_x$ have been considered:
\begin{itemize}
  \item The rate of three-jet events, where the jets are defined by the
        {\tt DURHAM} clustering algorithm with 
        the {\tt E} recombination scheme \cite{Durham}.
        As shown in Ref.~\cite{Delphibm1}, the optimal choice 
        of the resolution parameter $y_{\mathrm{cut}}$ for the
        mass determination is  $0.02$, because it
        represents a good compromise between the actual
        size of the mass effect to be measured  and the size of 
        backgrounds from other jet topologies. 
  
  \item The first and second moments of the event-shape variables
        thrust $T$, the $C$ parameter, the three-to-two jets 
        transition value $y_3$ and the 
        total and wide jet broadening, $B_T$ and $B_W$. The 
        definitions of the observables can be found, e.g., in 
        \cite{QCDmega, CataniBroad} and references therein. Moments have been
        chosen instead of distributions because of the better statistical
        accuracy with which the perturbative predictions can be
        evaluated by MC integration. Even higher moments would give
        smaller hadronization uncertainties. However, the NLO corrections
        typically get larger and the overall mass effects fall below
        the $3\%$ level.  
\end{itemize}   

\begin{table}[t]
  \caption{
    \protect\footnotesize
             Leading (LO) and next-to-leading (NLO) order
             contributions to $1 - R^{\mathrm{pert}}_{bd}$, for the
             running mass (run) and the pole mass (pol) schemes.
             The contributions are evaluated for a running (pole) mass
             of 3 (5) \gevcc. The strong coupling \asmz\ is set to 0.119.
             The values are given for the three-jet
             rate ({\tt DURHAM} scheme) and for the
             event-shape variables thrust $T$, $C$ parameter,
             the transition value $y_3$ for three to two jets ({\tt DURHAM}
             scheme), and the total  
             and wide jet broadenings ($B_T$ and $B_W$).
             The indices indicate the first or second moment of the event
             shape variable.
           }
  \vspace{0.3cm}
  \begin{center}
    \begin{tabular}{|c|r|r|r|r|}
      \hline
      \hline
      $O$ & LO (run)           & NLO (run)          & 
            LO (pol)           & NLO (pol)             \\
      \hline
      \hline
      \rule{0pt}{5.5mm}%
 $R_3      $&  $0.020\;\;$  &   $0.010\;\;\;\;$  &  $0.056\;\;$  &  $-0.008\;\;\;$   \\
      $T_1$ &  $0.036\;\;$  &   $0.019\;\;\;\;$  &  $0.076\;\;$  &  $-0.007\;\;\;$   \\
      $T_2$ &  $0.017\;\;$  &   $0.032\;\;\;\;$  &  $0.043\;\;$  &  $ 0.036\;\;\;$   \\
      $C_1$ &  $0.044\;\;$  &   $0.022\;\;\;\;$  &  $0.091\;\;$  &  $-0.011\;\;\;$   \\
      $C_2$ &  $0.021\;\;$  &   $0.039\;\;\;\;$  &  $0.052\;\;$  &  $ 0.043\;\;\;$   \\
  $y_{3_1}$ &  $0.032\;\;$  &   $0.007\;\;\;\;$  &  $0.071\;\;$  &  $-0.021\;\;\;$   \\
  $y_{3_2}$ &  $0.015\;\;$  &   $0.003\;\;\;\;$  &  $0.032\;\;$  &  $-0.007\;\;\;$   \\
  $B_{T_1}$ &  $0.117\;\;$  &   $0.006\;\;\;\;$  &  $0.188\;\;$  &  $-0.074\;\;\;$   \\
  $B_{T_2}$ &  $0.036\;\;$  &   $0.112\;\;\;\;$  &  $0.080\;\;$  &  $ 0.123\;\;\;$   \\
  $B_{W_1}$ &  $0.117\;\;$  &  $-0.085\;\;\;\;$  &  $0.188\;\;$  &  $-0.183\;\;\;$   \\
  $B_{W_2}$ &  $0.036\;\;$  &   $0.016\;\;\;\;$  &  $0.080\;\;$  &  $-0.013\;\;\;$   \\[2.5mm]
    \hline
    \hline
    \end{tabular}
  \end{center}
\label{lo-nlo:tab}
\end{table}

A list of the leading (LO) and 
next-to-leading order (NLO) contributions to $1 - R^{\mathrm{pert}}_{bd}$ is
given in Table \ref{lo-nlo:tab} for all variables, 
both for the running and the pole mass schemes. 
They have been evaluated for a $b$-quark mass of 3 \gevcc\ in the former and 5 \gevcc\
in the latter case. The computation of these terms is described 
in detail in Section \ref{theory}.
It is found that for some observables such as the jet
broadening variables the mass effect is rather large. However,
the NLO corrections can also be  sizeable, as in the case of the second moment
of thrust and the first moment of the wide jet broadening. For such 
variables it would definitely be necessary to also compute 
the NNLO contributions
in order to obtain a reliable perturbative prediction. 
Because of these observations in the following only those 
variables are considered for which in both schemes the NLO contribution
is clearly smaller than the LO term.
This requirement selects the three-jet rate ($R_3$),
the first moments of thrust ($T_1$), $C$ parameter ($C_1$), the three-to-two
jets transition value ($y_{3_1}$) and the total jet broadening ($B_{T_1}$),
as well as the second moment of the three-to-two
jets transition value ($y_{3_2}$) and the wide jet broadening ($B_{W_2}$).



\section{The ALEPH detector}
 \label{aleph}

The ALEPH detector is described in detail elsewhere
\cite{ALEPHdet1,ALEPHdet2}. Here only a description
of the tracking detectors is given, being the  relevant ones
for this analysis.
Briefly, at the core of the 
tracking system is a silicon strip vertex detector (VDET).
This has two layers, at average radii of 6.5 and 11.3 cm,
each providing measurements in both the $r$-$\phi$ and $r$-$z$
views. The spatial resolution for $r$-$\phi$ coordinates is
12 $\mu$m for normal incidence 
and varies between 12 and 22 $\mu$m for $z$ coordinates,
depending on the track polar angle. The angular coverage of 
the VDET is $|{ \cos\theta }| < 0.85$ for the inner layer and
$|{ \cos\theta }| < 0.69$ for the outer layer. The VDET lies
within a cylindrical drift chamber (ITC), which 
measures up to eight coordinates per track in the $r$-$\phi$
view, with a resolution of 150 $\mu$m. The ITC is in turn 
enclosed in a large time projection chamber (TPC), lying between
radii of 30 and 180 cm. The TPC provides up to 21 three-dimensional
coordinates per track, with resolutions in the $r$-$\phi$
and $r$-$z$ views of 180 $\mu$m and 500 $\mu$m, respectively. The 
three tracking detectors are  surrounded by a superconducting
solenoid producing a magnetic field of 1.5 T.

For charged tracks with two VDET coordinates, a transverse
momentum resolution of 
$\Delta p_T/p_T = 6 \times 10^{-4} p_T \oplus 0.005$
($p_T$ in GeV/$c$) is achieved. The three-dimensional 
impact parameter resolution
is $(25 + 95/p)\,\mu$m ($p$ in GeV/$c$).

Recently the LEP1 data set has been reprocessed using improved
reconstruction algorithms. In particular, a new VDET pattern
recognition algorithm allows groups of several nearby tracks,
which may share common hits, to be analyzed, in order to
find the hit-to-track assignment which minimizes the track fitting 
$\chi^2$  for the event as a whole. The improvement on the hit 
association efficiency is more than 2\%. Information from the
TPC wires, in addition to that obtained from the pads, is used
to improve the coordinate resolution by a factor of two in $z$,
and by 30\% in $r$-$\phi$ for low momentum tracks.

The tracking system is surrounded by electromagnetic and
hadronic calorimeters, and together they are used to measure
the neutral and charged energy flow.


\section{Data analysis}
 \label{data}

\subsection{Event selection}
 \label{eventsel}

In this analysis data taken at the peak of the Z resonance
from 1991 to 1995 are used.
A standard hadronic event selection \cite{QCDmega} is applied,
which is based on charged particles. A
cut $|{ \cos\theta_{T} }| < 0.7$ is imposed, where
$\theta_{T}$ is the polar angle of the thrust axis, computed
from all charged and neutral particles as obtained from the 
energy-flow algorithm \cite{EFLOW}. This requirement ensures that
the events are well contained within the VDET acceptance.
According to the MC simulation the event selection is 61.7\% efficient.  
Non-hadronic background, which is dominated by $\tau^+\tau^-$ events, 
amounts to 0.3\% of this sample. After the selection, 
a sample of 2.3 million
hadronic events remains for further analysis. 

The analysis also uses 8.8 million simulated hadronic events 
produced with a generator based on the {\tt JETSET 7.4} parton shower
model \cite{Jetset}. The production rates, decay modes and lifetimes
of heavy hadrons are adjusted to agree with recent measurements,
while heavy quarks are fragmented using the Peterson et al.\ model 
\cite{Peterson}. Detector effects are simulated using the {\tt GEANT} package
\cite{GEANT}. The MC events are reweighted in order to 
reproduce the measured values for the gluon splitting rates
into $c\bar{c}$ and $b\bar{b}$  pairs, which are
$g_{c\bar{c}} = 0.0319 \pm 0.0046$ and 
$g_{b\bar{b}} = 0.00251 \pm 0.00063$ \cite{LEPHF}.

The observables 
described in Section \ref{method} are computed using only
charged tracks. This choice is preferred over taking
both charged and neutral energy-flow objects, as 
for this analysis it allows the
reduction of the detector related systematic uncertainties 
to below the 1\% level, without reducing the statistics.

For the computation of the three-jet
rate, an additional cut on the minimum energy of a jet of 5 GeV
is applied, which removes about 5\% of all three-jet events. 
The three-jet rates are
17.7\% (18.2\%) in data (MC) for $b$-tagged events, and
19.3\% (20.4\%)  in the inclusive sample. The excess of three-jet
events in the MC simulation had been observed previously \cite{QCDmega}.


\begin{figure}[t]
  \begin{center}
    \includegraphics[width=11cm]{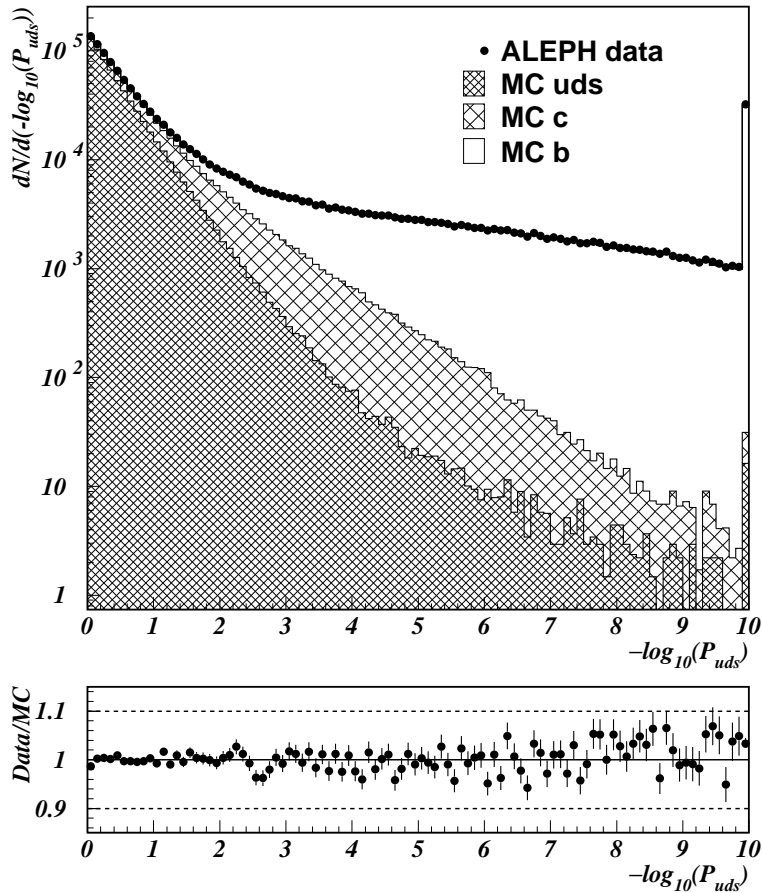}
    \caption{\footnotesize Distribution of the $b$-tag variable for data and
    Monte Carlo $b$, $c$ and $uds$ events. The Monte Carlo has
    been normalized to the same number of events as the data. The last bin
    includes overflow entries. The insert at the bottom shows the
    ratio of data over Monte Carlo. \label{fig:Puds}}
  \end{center}
\end{figure}

\subsection{The \boldmath$b$\unboldmath-tag algorithm}
 \label{btag}

The $b$-tag algorithm, i.e., the criteria used to select $b$ events,
follows rather closely the approach described in detail in
Ref.~\cite{Rb}. Briefly, the presence of
$b$ hadrons is detected using a tag based on the long
$b$ hadron lifetime and the precision of the VDET.
In contrast to \cite{Rb},
where the tag is applied separately to both hemispheres of the event,
here the algorithm uses all tracks of the event.

The actual  selection of $b$ events is obtained from a cut on the
distribution of the confidence level $P_{uds}$ 
that all tracks of the event come
from the main vertex. This distribution as measured in data and in
MC is shown in Fig.~\ref{fig:Puds}, 
together with the expected contributions from
different flavours.
Very good agreement between data and MC is observed. The requirement
$-\log_{10} P_{uds} > 2.2$ selects $b$ events with an efficiency
of $\epsilon_b = 80.5\%$ and a purity of ${\cal P}_{b} = 83.1\%$. The 
backgrounds amount to ${\cal P}_{c} = 13.5\%$ and ${\cal P}_{l} = 3.4\%$.

With this selection, about $488\,000$ $b$-tagged events are found in the data
before the three-jet requirement.
Out of about $435\,000$ three-jet events, $86\,000$
are tagged as $b\bar{b}$.


\subsection{Detector and tagging corrections}
 \label{detcorr}

The observables must be corrected for detector effects, such as
acceptance and resolution. This is achieved by computing
the observables before and after detector simulation,
imposing the same track and event selection cuts as 
for the data, except for the flavour tag. The ratios
of the observables are computed for each true flavour, i.e.,
the quark flavour of the primary decay products of the Z boson.
When computing the observables at the true hadron level before
detector simulation, initial state radiation is 
turned off. In the case of the three-jet rate the detector corrections 
are $D_b = 0.909$ and $D_{l}=0.843$, with a statistical
uncertainty at the few per mille level.

The flavour tag can introduce a bias on the measured
observables. The bias is estimated by computing
the observable before and after applying the tag to the MC sample 
which passed the event selection cuts,
for every flavour. As an example, $T_{b} = 0.894$ is found
for the three-jet rate. Taking the product of ratios of detector
corrections, tagging biases and purities the overall detector
corrections amount to $D_{b/l}\,T_{b}\,{\cal P}_{b}=0.801$,
$D_{c/l}\,T_{c}\,{\cal P}_{c}=0.112$ and 
$T_{l}\,{\cal P}_{l}=0.045$, again within a
statistical accuracy of a few per mille.


\subsection{Hadronization corrections}
 \label{hadcorr}

The perturbative predictions are corrected for hadronization effects
by computing the relevant observables at parton and at hadron level,
including final state photon radiation off quarks.
Several Monte Carlo models based on the parton shower approach
plus subsequent string or cluster fragmentation are employed for
this purpose. For the nominal analysis, the same generator as for the
full simulation is used, which is {\tt HVFL} \cite{HVFL}.  
As already mentioned in Section \ref{eventsel}, this generator is 
based on the {\tt JETSET 7.4} parton shower model plus string 
fragmentation, and in addition the decay modes and lifetimes
of heavy hadrons are adjusted to agree with recent measurements.
Heavy quarks are fragmented using the Peterson et al.\ model.
The fragmentation parameters for this generator are determined from a 
global fit to hadronic Z decay data as described in \cite{QCDmega},
using the present knowledge of heavy quark physics.
The modelling of heavy quark physics represents an important
part of the systematic uncertainty of the result.

In the {\tt JETSET} parton shower model mass effects are introduced by
kinematic constraints to the phase space at each parton branching
in the shower evolution as well as by the matching to the
leading order matrix element at the first parton branching. 
The value of $M_b$ (pole mass) is set to $4.75\,\gevcc$. 
Thirty million events without detector
simulation were  generated and analyzed.
Such a large sample is necessary in order to reduce the statistical
error of the hadronization corrections below the one per mille level.
These corrections were computed for each flavour individually.

\begin{table}[t]
    \caption{
      \protect\footnotesize 
             Results for $R^{\mathrm{pert}}_{bd}$ with statistical
             errors, including the uncertainty from the MC statistics.
             The results are obtained with the hadronization 
             corrections $H_{b/l}$ as predicted by {\tt HVFL}. Also given are
             the corrections $H^{nod}_{b/l}$, computed from hadrons
             directly originating from the string, before any decays
             (\textit{nod}=no decays).
            }
  \vspace{0.3cm}
  \begin{center}
    \begin{tabular}{|c|c|c|c|}
      \hline
      \hline
      $O$ & $R^{\mathrm{pert}}_{bd}$ & $\;H_{b/l}\;$ & $\;H^{nod}_{b/l}\;$ \\
      \hline
      \hline
      \rule{0pt}{5.5mm}%
%
             $R_3$ &  $0.974\pm0.005$& 0.980  &0.987  \\
             $T_1$ &  $0.910\pm0.002$& 1.134  &1.003  \\
             $C_1$ &  $0.894\pm0.002$& 1.169  &1.009  \\
         $y_{3_1}$ &  $0.955\pm0.005$& 1.023  &0.987  \\
         $y_{3_2}$ &  $0.979\pm0.010$& 0.984  &0.984  \\
         $B_{T_1}$ &  $0.831\pm0.001$& 1.306  &1.025  \\
         $B_{W_2}$ &  $0.929\pm0.003$& 1.088  &0.986  \\[2.5mm]
%
    \hline
    \hline
    \end{tabular}     
  \end{center}
\label{statres:tab}
\end{table}

\begin{figure}[t]
  \begin{center}
    \includegraphics[width=10cm]{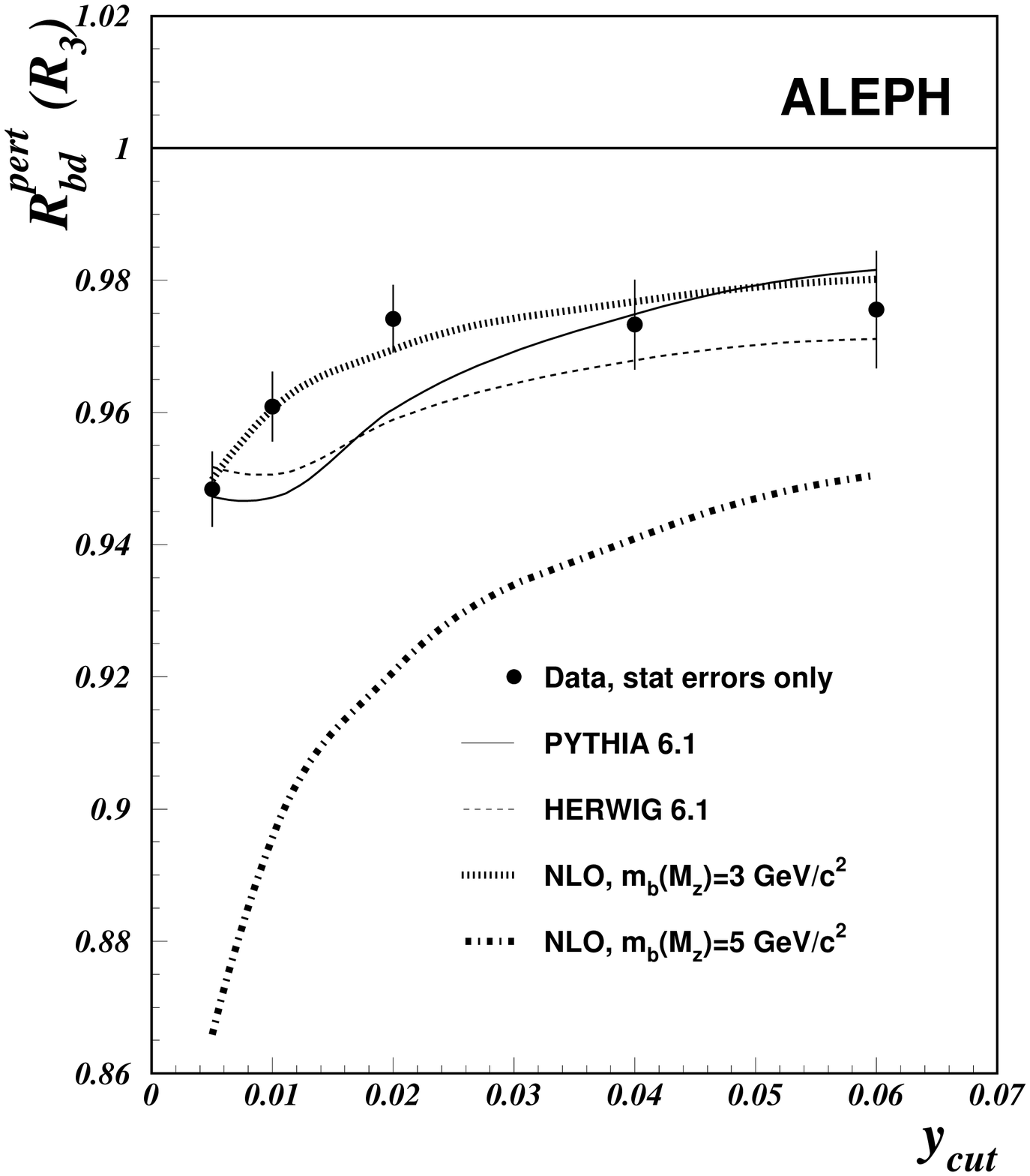}
    \caption{\footnotesize
    Comparison of the $y_{\mathrm{cut}}$ dependence of the measured
    ratio $R^{\mathrm{pert}}_{bd}$ for the three-jet rate 
    to the predictions of parton shower models
    as well as next-to-leading order (NLO) perturbative QCD for two
    different values of the 
    $b$-quark mass in the $\overline{\mathrm{MS}}$ scheme.
    The measurement is obtained using {\tt HVFL} for the hadronization
    corrections. The errors are statistical only. \label{fig:r3bd_ycut}}
  \end{center}
\end{figure}

The ratios of hadronization corrections, $H_{b/l}$,
are listed in Table \ref{statres:tab}. The 
corrections are rather sizeable for almost all the observables; in most
cases they are of the same size as or larger than the expected mass effect.
Only the three-jet rate and the first two moments of the $y_3$
distribution have corrections at the 2\% level or below. 
For all the other event-shape variables the corrections are of 
the order of 10\% or even larger. 
It has been found that the deviation from unity
is almost entirely due to B hadron decays, which change the distributions
mainly in the two-jet region. As can be observed from Table \ref{statres:tab},
the same corrections, computed taking only hadrons which stem directly
from the string before any decay, come close to unity within one or
two percent. This is in agreement with expectations from recent 
calculations of nonperturbative power-law corrections to moments
of event-shape distributions for massive quarks \cite{Trocsanyi:2000ta}.
The strong dependence on the decays of heavy quarks is 
taken into account when estimating systematic uncertainties in the following.

\subsection{Measurement of \boldmath$R^{\mathrm{pert}}_{bd}$\unboldmath}
 \label{rmeas}

The correction procedure defined in Eqn.~\ref{correction} has been carried
out individually for every data taking year. The hadronization corrections
are taken from the {\tt HVFL} generator and do not depend on the year.
Good agreement between the values of the corrected  
observable $R^{\mathrm{pert}}_{bl}$ in the different years is found, and therefore
the final value is obtained from the average over years, weighted according to
the statistical errors. 
The $\chi^2$ confidence level for the combination is 50\%.
In order to obtain the final values for $R^{\mathrm{pert}}_{bd}$, the
contributions from anomalous triangle diagrams are first subtracted from the
measured ratio $R^{\mathrm{pert}}_{bl}$, and then the error due to the finite
MC statistics for the evaluation of purities and detector and tagging corrections
is added to the statistical error of the data. This statistical
error from MC is obtained by evaluating the scatter of results when repeating
the analysis with a large number of MC subsamples of smaller size, and
then extrapolating the standard deviation thus found to the actual 
size of the original MC sample.  The results are listed in
Table \ref{statres:tab}. 

The dependence on the resolution parameter $y_{\mathrm{cut}}$
of the perturbative ratio $R^{\mathrm{pert}}_{bd}$ for the three-jet rate
is indicated in Fig.~\ref{fig:r3bd_ycut}.
The data are compared to the predictions of the parton shower
models of {\tt PYTHIA 6.1}, which is based on {\tt JETSET},
and {\tt HERWIG 6.1} \cite{Herwig}. The MC models
are in reasonable agreement with
the measurement at large $y_{\mathrm{cut}}$ values, where the statistical
uncertainty is large, however. At intermediate resolution
parameters these models predict a lower ratio than observed.
In the lowest $y_{\mathrm{cut}}$ region, where the resolution parameter
approaches the scale of the $b$-mass, a turn-over is observed,
which might be interpreted as a short-coming of the approximation
for mass effects as implemented in these models.

In addition, in Fig.~\ref{fig:r3bd_ycut} the next-to-leading order perturbative
QCD predictions for two different values of the running $b$-quark mass
in the $\overline{\mathrm{MS}}$ scheme are given.
The data clearly favour a mass close to 3 \gevcc.


\subsection{Systematic uncertainties}
 \label{systematics}


\subsubsection{Detector and physics modelling}

Systematic uncertainties can arise
from imperfections of the implementation of the physics processes
as well as the description of the detector performance.
An important physics parameter for the correct description of three-jet rates
as well as tagging purities is the gluon splitting rate into heavy quark
pairs ($b\bar{b}$, $c\bar{c}$). The MC was 
reweighted in order to reproduce the most recent measured values 
and the experimental uncertainties were
propagated to a systematic error on $R^{\mathrm{pert}}_{bd}$.

The MC predictions for the background efficiencies depend on the mean number
of VDET coordinates per track and on the assumed impact parameter resolution.
It is observed that in the MC simulation more tracks are accepted for use
by the $b$-tag algorithm than in the data, and that the resolution function in
impact parameter significance is not perfectly modelled. In order to 
improve the agreement between data and MC, tracks in the MC are deleted
randomly, and a smearing algorithm is applied to the impact parameter 
significance of some of the MC tracks. A detailed description of this
procedure is given in Ref.~\cite{Rb}. The systematic uncertainty related
to this procedure is estimated by repeating the whole analysis once
without the track deletion, and once without any smearing. 
Half of the observed deviations in the final result are taken as uncorrelated
errors and added  in quadrature.

In the case of the three-jet rate additional cuts
on the jet quality were studied.  
Removing the cut on the minimum jet energy results in a relative shift of $-0.3\%$
for $R^{\mathrm{pert}}_{bd}$.
As a further check of the quality of the MC description of jet quantities,
the requirement of at least three charged tracks per jet is applied.
This causes a relative change in the result of $-0.3\%$. 
A relative systematic uncertainty of $0.3\%$ related to the modelling of 
three-jet events is therefore assumed.

The relative uncertainties due to the physics and detector modelling  as
described above are listed in Table \ref{systres:tab}.
Adding all these uncertainties
in quadrature results in a final experimental systematic error (exp) 
as quoted in Table \ref{fullres:tab} for all the observables under
consideration.


\subsubsection{Hadronization}
  \label{hadsys}

The uncertainties from the modelling of the hadronization are typically
evaluated by computing the hadronization corrections with different MC 
generators. However, if these uncertainties are to be meaningful, it
must be verified that the various models give a good overall
description of hadronic Z decay data, and in particular of quantities
relevant to the analysis. 

As has been shown in Section \ref{hadcorr}, the B hadron decays have
a large impact on the size of the hadronization corrections.
Because in the tuning of the MC hadronic final states are analyzed
after all decays, differences in the description of hadron decays can
lead to differences in the tuned fragmentation parameters.
In fact,
comparing the predictions of the standard {\tt PYTHIA 6.1} and the
{\tt HVFL} generator, both based on the {\tt JETSET} parton shower and string fragmentation
and tuned to ALEPH data, differences in the
fragmentation parameters are found, which translate to a variation in the
hadronization corrections, even before considering decays.
In order to assess an uncertainty related to the string fragmentation 
parameters and subsequent decays,
the variation in the final result when using {\tt HVFL} or {\tt PYTHIA} for the
hadronization corrections is taken as a systematic uncertainty.

Another quantity relevant for this analysis 
is the $b$ fragmentation function, which describes the
energy fraction transferred to the B hadrons during the fragmentation
process. In particular, the lower energy tail of 
the fragmentation function is important for three-jet events.
Recently, new measurements of the $b$ fragmentation function
have been performed \cite{ALEPH:bfrag, SLD:bfrag}. 
Both {\tt HVFL} and {\tt PYTHIA} use the Peterson model in their standard setup.
In order to test the sensitivity to the $b$ fragmentation function,
the measured distribution \cite{ALEPH:bfrag} was included
in a global tuning of the {\tt PYTHIA} fragmentation parameters, and a new
set of hadronization corrections was computed with those
parameters. 
In Ref.~\cite{SLD:bfrag} it is shown that the Lund fragmentation scheme
combined with a model by Bowler \cite{Jetset, Bowler} gives a good
description of the $b$ fragmentation function. 
The {\tt PYTHIA} MC was therefore tuned to the data again, but now using the Bowler
fragmentation scheme instead of the Peterson model, and new
hadronization corrections were derived accordingly. 
The results obtained with the retuned parameter set and
the Bowler fragmentation function were then compared to the 
standard {\tt PYTHIA} predictions.
The maximum deviation is taken as an estimator of the uncertainty
related to the $b$ fragmentation function.

The {\tt HERWIG 6.1} MC generator was used for the study of a 
fragmentation model different from the string approach. As discussed
above, differences in the description of hadron decays can effectively
lead to differences in the pure fragmentation parameters, after tuning.
In order to reduce the sensitivity to this effect, and to study
purely the difference between string and cluster fragmentation, 
the variation in the hadronization corrections for light quarks ($uds$)
only has been propagated into an uncertainty  on the final result.

The three components of uncertainty from the hadronization process
as described above are listed in Table \ref{systres:tab}.
They are then added in quadrature
and quoted as the hadronization uncertainty (had) in Table \ref{fullres:tab}.
It is the dominant uncertainty for all variables.

\begin{table}[t]
    \caption{
      \protect\footnotesize 
             Variations (in percent) of $R^{\mathrm{pert}}_{bd}$
             with respect to the nominal value for all systematic studies performed
             (PY={\tt PYTHIA}). The first block indicates the components of the
             statistical uncertainty from data and MC, 
             in the second block the experimental uncertainties due to the
             physics and detector modelling are given, 
             the third block
             lists the contributions to the systematic uncertainty due to
             the modelling of hadronization, and the last block gives the
             results of further systematic checks. 
            }
  \vspace{0.3cm}
  \begin{center}
    \begin{tabular}{|l|r|r|r|r|r|r|r|}
      \hline
      \hline
      \rule{0pt}{4.5mm}%
   & \multicolumn{1}{|c|}{$R_3$}       & 
     \multicolumn{1}{|c|}{$T_1$}       & 
     \multicolumn{1}{|c|}{$C_1$}       & 
     \multicolumn{1}{|c|}{$y_{3_1}$}   & 
     \multicolumn{1}{|c|}{$y_{3_2}$}   & 
     \multicolumn{1}{|c|}{$B_{T_1}$}   & 
     \multicolumn{1}{|c|}{$B_{W_2}$} \\[0.3mm] 
      \hline
      \hline
   & \multicolumn{7}{|c|}{statistical uncertainties} \\ 
      \hline
      \rule{0pt}{5.5mm}%
   Stat.\ error Data   &
      $ 0.44\; $&$ 0.19 $&$ 0.15 $&$ 0.43 $&$ 0.90 $&$ 0.11 $&$ 0.28  $\\
   Stat.\ error MC &
      $ 0.28\; $&$ 0.14 $&$ 0.11 $&$ 0.28 $&$ 0.55 $&$ 0.07 $&$ 0.20  $\\
      \hline
      \hline
     & \multicolumn{7}{|c|}{experimental uncertainties} \\ 
   \hline
      \rule{0pt}{5.5mm}%
   No hit smearing &
      $-0.47\; $&$-0.26 $&$-0.22 $&$-0.42 $&$-0.50 $&$-0.14 $&$-0.36  $\\
   No track deletion & 
      $ 0.71\; $&$ 0.38 $&$ 0.29 $&$ 0.78 $&$ 1.10 $&$ 0.17 $&$ 0.57  $\\
   $E_{\rm ch}/{\rm Jet} \ge 0$ &
      $-0.29\; $&$ 0.00 $&$ 0.00 $&$ 0.00 $&$ 0.00 $&$ 0.00 $&$ 0.00  $\\
   $N_{\rm ch}/{\rm Jet} \ge 3$ &
      $-0.28\; $&$ 0.00 $&$ 0.00 $&$ 0.00 $&$ 0.00 $&$ 0.00 $&$ 0.00  $\\
   Gluon splitting &
      $-0.14\; $&$ 0.10 $&$ 0.16 $&$-0.18 $&$-0.44 $&$ 0.25 $&$ -0.03 $\\
      \hline
      \hline
     & \multicolumn{7}{|c|}{hadronization uncertainties} \\ 
   \hline
      \rule{0pt}{5.5mm}%
   Fragm.\ parameters & &       &       &           &           &           & \\
   and B decays & 
      $-1.00\; $&$ 0.93 $&$ 1.42 $&$-0.58 $&$-0.65 $&$ 3.20 $&$-0.16  $\\
   Fragm.\ model &
      $ 0.64\; $&$-3.82 $&$-4.33 $&$ 0.67 $&$ 1.56 $&$ 4.42 $&$-0.89  $\\
   $b$ fragm.\ function &
      $ 0.27\; $&$-0.33 $&$-0.41 $&$ 0.14 $&$ 0.42 $&$-0.56 $&$-0.22  $\\
      \hline
      \hline
     & \multicolumn{7}{|c|}{additional systematic checks} \\ 
   \hline
      \rule{0pt}{5.5mm}%
   $c$-quark mass effect &
      $-0.03\; $&$-0.12 $&$-0.15 $&$-0.09 $&$-0.02 $&$-0.34 $&$-0.11  $\\
   PY 6.1, $\mathcal{O}(\as)$ massless &
      $ 0.40\; $&$-2.03 $&$-2.32 $&$ 0.14 $&$ 0.60 $&$-2.90 $&$-0.21  $\\
   PY 6.1, $Q_0=4$ GeV &
      $-0.86\; $&$ 1.10 $&$ 1.47 $&$-0.82 $&$-0.66 $&$ 3.69 $&$ 0.27  $\\
   $-\log_{10} P_{uds} > 1.2$ & 
      $-0.25\; $&$-0.57 $&$-0.71 $&$-0.05 $&$ 0.13 $&$-0.82 $&$-0.45  $\\
   $-\log_{10} P_{uds} > 4.0$ &
      $ 0.27\; $&$ 0.82 $&$ 0.77 $&$ 0.69 $&$ 0.67 $&$ 0.67 $&$ 0.56  $\\
   $R^{\mathrm{meas}}_{bl}$ &
      $ 0.03\; $&$-0.04 $&$-0.11 $&$ 0.14 $&$ 0.30 $&$-0.14 $&$-0.09  $\\[2.5mm]
   \hline
   \hline   
    \end{tabular}     
  \end{center}
\label{systres:tab}
\end{table}

\begin{table}[t]
    \caption{
      \protect\footnotesize 
             Results for $R^{\mathrm{pert}}_{bd}$ with statistical
             (stat), experimental (exp) and hadronization (had) uncertainties.
            }
  \vspace{0.3cm}
  \begin{center}
    \begin{tabular}{|c|cccc|}
      \hline
      \hline
      $O$ & $\;\;\;R^{\mathrm{pert}}_{bd}\;\; $ & $\;\;\pm$(stat)$\;$ & 
                                         $\;\;\pm$(exp) $\;$ & 
                                         $\;\;\pm$(had) $\;$\\
      \hline
      \hline
      \rule{0pt}{5.5mm}%
             $R_3$ & 0.974 & 0.005 & 0.005 & 0.012 \\
             $T_1$ & 0.910 & 0.002 & 0.004 & 0.036 \\
             $C_1$ & 0.894 & 0.002 & 0.004 & 0.041 \\ 
         $y_{3_1}$ & 0.955 & 0.005 & 0.005 & 0.009 \\ 
         $y_{3_2}$ & 0.979 & 0.010 & 0.007 & 0.017 \\ 
         $B_{T_1}$ & 0.831 & 0.001 & 0.005 & 0.046 \\ 
         $B_{W_2}$ & 0.929 & 0.003 & 0.003 & 0.009 \\[2.5mm]
    \hline
    \hline
    \end{tabular}     
  \end{center}
\label{fullres:tab}
\end{table}


\subsubsection{Additional systematic checks}
 \label{xchecks}

A number of additional systematic checks have been
performed in order to evaluate the stability of the results.
They are listed in Table \ref{systres:tab}.

In Eqn.~\ref{correction} the world average values for $R_b$ and
$R_c$ enter. Changing them within their errors has a negligible 
impact on the results. 

An additional factor which contributes to Eqn.~\ref{correction}
is $R^{\mathrm{pert}}_{cl}$, which is the ratio of the observable
at parton level for $c$ and light quark events. Because of
the smaller $c$-quark mass this ratio is set to unity in the default
analysis. Considering a
$c$-quark mass of $m_c = 1.4\,\gevcc$
leads to very small variations. Only in the case of the total
jet broadening is the observed shift larger than the statistical
uncertainty; in that case the shift is therefore
added in quadrature to the experimental uncertainty.

A possible bias 
from the $b$-quark mass in the parton level simulation is estimated
by switching off the correction to the massive matrix element for the
first branching of a massive quark in the parton shower.
For most of the variables, and in particular for the three-jet rate,
this results in small variations.
As a further cross check, a variant of the {\tt JETSET} fragmentation model
was studied in which the parton shower cutoff parameter $Q_0$
is increased to 4 GeV. This setting leads to a parton multiplicity
of 4 on average, which is close to the multiplicity used in the
analytical next-to-leading order calculations. However, it represents
a drastic change in the model, and such large cutoff values do not
agree with the basic ideas of the parton shower approach. Nevertheless,
recomputing $R^{\mathrm{pert}}_{bd}$ with the hadronization corrections
obtained from this model leads to variations that are well within
the fragmentation uncertainty quoted above.

In order to check the stability of the result with respect to the
chosen $b$-tag working point, the cut on the output of the
$b$-tag algorithm was varied over a wide range, which resulted
in changes of the purity ${\cal P}_{b}$ from 64\% to 95\%. 
For most of the variables, in particular the three-jet rate,
the observed deviations in the final result
are within the tracking uncertainty as described above. However,
in those cases where the deviation exceeds this uncertainty,
half of the maximum deviation under variation of the $b$-tag
working point is taken as a systematic uncertainty due to the tracking.

An attempt was made to verify with data the purity
${\cal P}_{b}$  determined from MC.
The approach is based on the double tag 
method as described in \cite{Rb}. It is possible to extract
the hemisphere tagging efficiencies, which then can be related to a
global event tagging efficiency if the hemisphere correlations are known. 
The same working point as in the nominal analysis was chosen. 
In a first check, $R_b$ and the hemisphere $b$-tag efficiency $\epsilon^h_b$
are measured from the data, whereas all the background efficiencies
and the hemisphere correlations are taken from MC. 
The results are $R_b=0.2190\pm0.0013(\mathrm{stat})$, 
which is consistent with the world average
value, and $\epsilon^h_b=0.5380\pm0.0024(\mathrm{stat})$. Next  $R_b$ is fixed
to the world average value, and $\epsilon^h_b$ as well as the
hemisphere correlation $\rho_b$ are measured. The result can be translated
to the global efficiency by the relation $\epsilon_b = 2\,\epsilon^h_b -
(\epsilon^h_b)^2\,(1+\rho_b)$, and then the purity is obtained
as  
\be
   {\cal P}_{b} = \frac{R_b\,\epsilon_b}{R_b\,\epsilon_b\, +\,
                                          R_c\,\epsilon_c\, +\,
                                  (1-R_b-R_c)\,\epsilon_{uds}} \quad. 
\ee
The result is ${\cal P}_{b} = 0.8310\pm0.0007(\mathrm{stat})$, which is in
very good  agreement with the value extracted from the MC, 
${\cal P}^{\mathrm{MC}}_{b} = 0.8305$.

In order to test the overall correction method, an independent
set of MC events was analyzed in the same way as the data.
Within statistical errors, the extracted 
$R^{\mathrm{pert}}_{bl}$ values are in agreement with the ``true''
values as found at the parton level of the MC prediction. 

Finally, the whole measurement is repeated using the method
of Ref.~\cite{Delphibm1}, where $R^{\mathrm{meas}}_{bl}$ is 
measured instead of $R^{\mathrm{meas}}_{bq}$. Both methods have the
same statistical accuracy, but the latter is more simple from the
point of view of systematic uncertainties. 
For the purpose of measuring $R^{\mathrm{meas}}_{bl}$
a $uds$ tag must also be applied. This is achieved by the requirement
$-\log_{10} P_{uds} < 0.5$, which selects light quark
events with an efficiency of 66\% and a purity of ${\cal P}_{l} =88$\%.
The correction procedure and the correction factors needed are
different; nevertheless the results obtained differ 
from the nominal ones by at most $0.3\%$.



\subsection{Results for \boldmath$R^{\mathrm{pert}}_{bd}$\unboldmath}
 \label{results}

As stated in Section \ref{hadcorr}, the nominal results are obtained
when using {\tt HVFL} for the hadronization corrections.
The results for all observables can be found in Table \ref{fullres:tab},
together with statistical and systematic uncertainties.
The result for the three-jet rate is in agreement 
with that of Ref.~\cite{Delphibm1}.
Most of the event-shape variables have a better statistical accuracy
than the three-jet rate. However, the hadronization uncertainties
are rather large. 

For the extraction of the $b$-quark mass it is required
that the uncertainty on the measured ratio
be at the 1\% level, and that at the same time the size of the
hadronization correction (Table~\ref{statres:tab})
be smaller than the actually measured
mass effect, given by the deviation of $R^{\mathrm{pert}}_{bd}$ from unity.
These requirements leave only the first moment of the $y_3$ distribution
and the three-jet rate for further investigation.


\section{Determination of the \boldmath$b$\unboldmath-quark mass}
 \label{massextraction}

\subsection{Theoretical predictions}
 \label{theory}

The NLO prediction for $R^{\mathrm{pert}}_{bd}$ as a function of 
the observable $O$ and the $b$-quark mass can be cast in the following form:
\be
  \label{nlogeneral}
      R^{\mathrm{pert}}_{bd}(O) = 1 + 
       r_b(\mu) \left[ b_0(r_b(\mu),O) + \frac{\as(\mu)}{2\pi} b_1(r_b(\mu),O) \right] \; = \; 1 + K(r_b(\mu))
       \quad ,
\ee
where $r_b(\mu) = m_b^2(\mu)/M_{\mathrm Z}^2$, and $m_b(\mu)$ is the running $b$-quark mass
as defined in the $\overline{\mathrm{MS}}$ renormalization scheme at the 
renormalization scale $\mu$. 
The ratio can also be expressed in terms of the pole mass
$M_b$, 
\be
  \label{pole}
      R^{\mathrm{pert}}_{bd}(O) = 1 + 
       r^P_b \left[ b^P_0(r^P_b,O) + \frac{\as(\mu)}{2\pi} b^P_1(r^P_b,O) \right]
       \quad ,
\ee
with $r^P_b = M_b^2/M_{\mathrm Z}^2$. The two predictions are equivalent at this order. The 
coefficient functions $b_{0,1}$ for the two schemes can be related to each other
by expressing the pole mass in terms of the running mass, i.e.,
\be
 \label{poletorun}
    r^P_b = r_b(\mu) \left[ 1 + \frac{\as(\mu)}{2\pi}
            \left( \frac{16}{3} - 4 \ln r_b(\mu) + 4 \ln\frac{\mu^2}{M_{\mathrm Z}^2} \right)
                     \right] \quad .
\ee
The coefficient functions for the three-jet rate were computed in 
Ref.~\cite{rodrigo1}. For all the other variables they were obtained
using the MC generators ZBB4 \cite{Nasonbmass, Nasonprivate} and EVENT \cite{Event}. ZBB4
allows for the integration of
the fully differential NLO matrix elements including
mass effects, whereas EVENT contains the massless expressions \cite{ERT}. The latter
has been extensively employed at LEP for the calculations of the NLO predictions
of event shapes used for the measurements of the strong coupling constant.
The differential cross sections for any infrared and collinear safe observable 
$O$ for either $b$ or $d$ primary quarks in 
$\mathrm{e}^+\mathrm{e}^-$ annihilation
can be written as
\be
 \label{diffxsec}
    \frac{1}{\sigma_{b,d}^{tot}} \frac{d \sigma_{b,d}}{d O} = 
    \frac{\alpha_s}{2\pi}\, A_{b,d}(O) + 
    \left(\frac{\alpha_s}{2\pi}\right)^2 B_{b,d}(O) \quad .
\ee  
Here $\sigma_{b,d}^{tot}$ is the total cross section for either $b$ or $d$
quark production. The functions $A_b, B_b$ are computed with ZBB4
for fixed values of the $b$-quark mass in the pole mass scheme, whereas
$A_d$ and $B_d$ are obtained from EVENT. Taking the ratio of the cross sections
for $b$ and $d$ quarks and expanding
up to NLO, the following relationships can be found:
\be
 \label{rela1}
   r^P_b b^P_0 = \frac{A_b}{A_d} - 1  \quad ,
\ee
\be
 \label{rela2}
   r^P_b b^P_1 = \frac{B_b A_d - B_d A_b}{A_d^2} \quad .
\ee
In order to get the functional dependence of $b^P_{0,1}$ on the
$b$-quark mass, the coefficients $A_b$ and $B_b$ were first computed in very
high statistics runs for values of $M_b = 1,2,3,4,5,6\, \gevcc$.  The
expressions (\ref{rela1}) and (\ref{rela2}) were then fitted over the relevant
range of 3 to 6 \gevcc\ using parametrisations of the type
$c_1 + c_2 r^P_b + c_3 \ln r^P_b$.
This functional form follows closely that used in \cite{rodrigo1}.
It has been checked that different parametrisations lead to very small
changes in the extracted mass values.

\begin{figure}[t]
  \begin{center}
    \begin{tabular}{cc}
      \includegraphics[width=8cm]{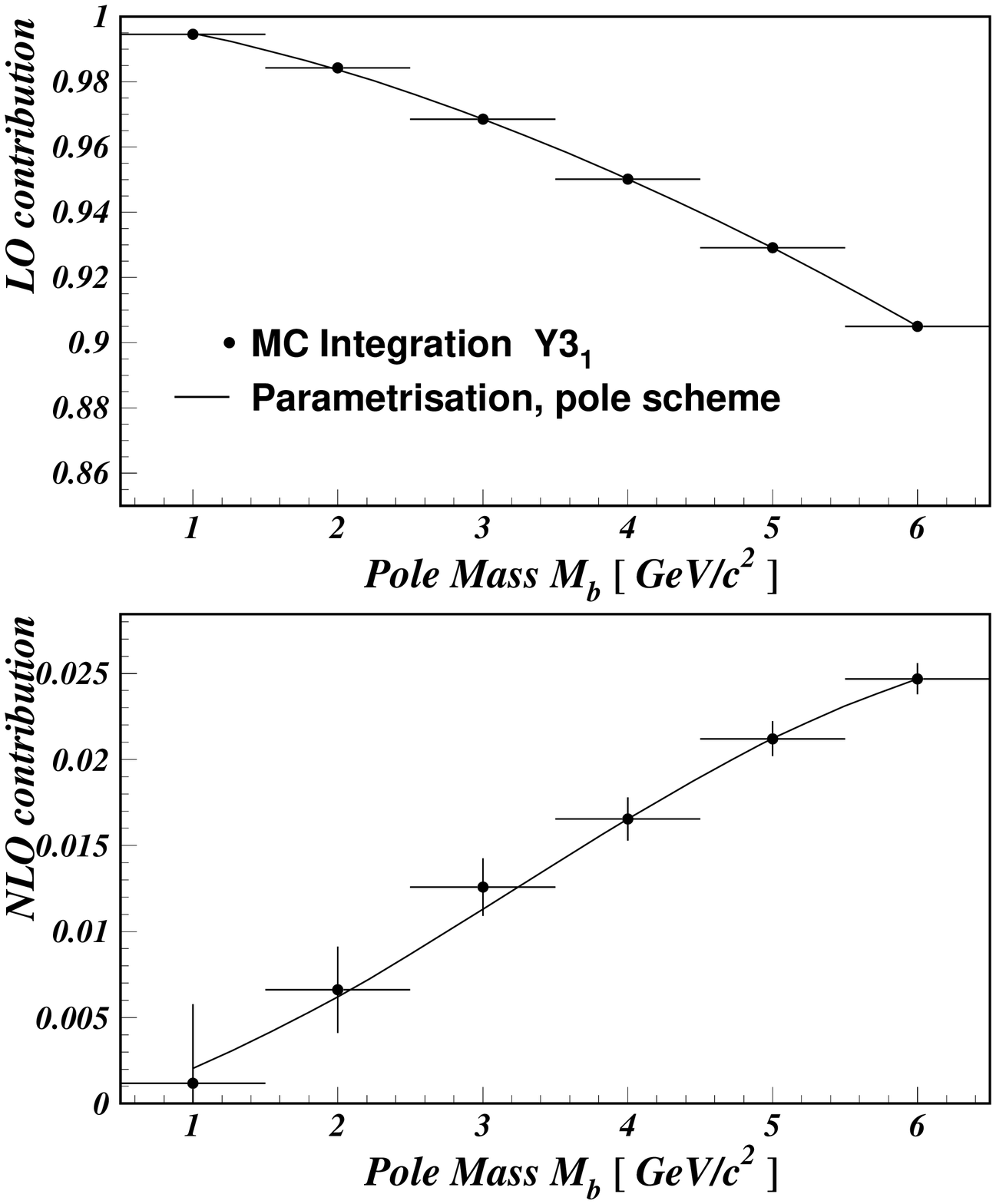} &
      \includegraphics[width=8cm]{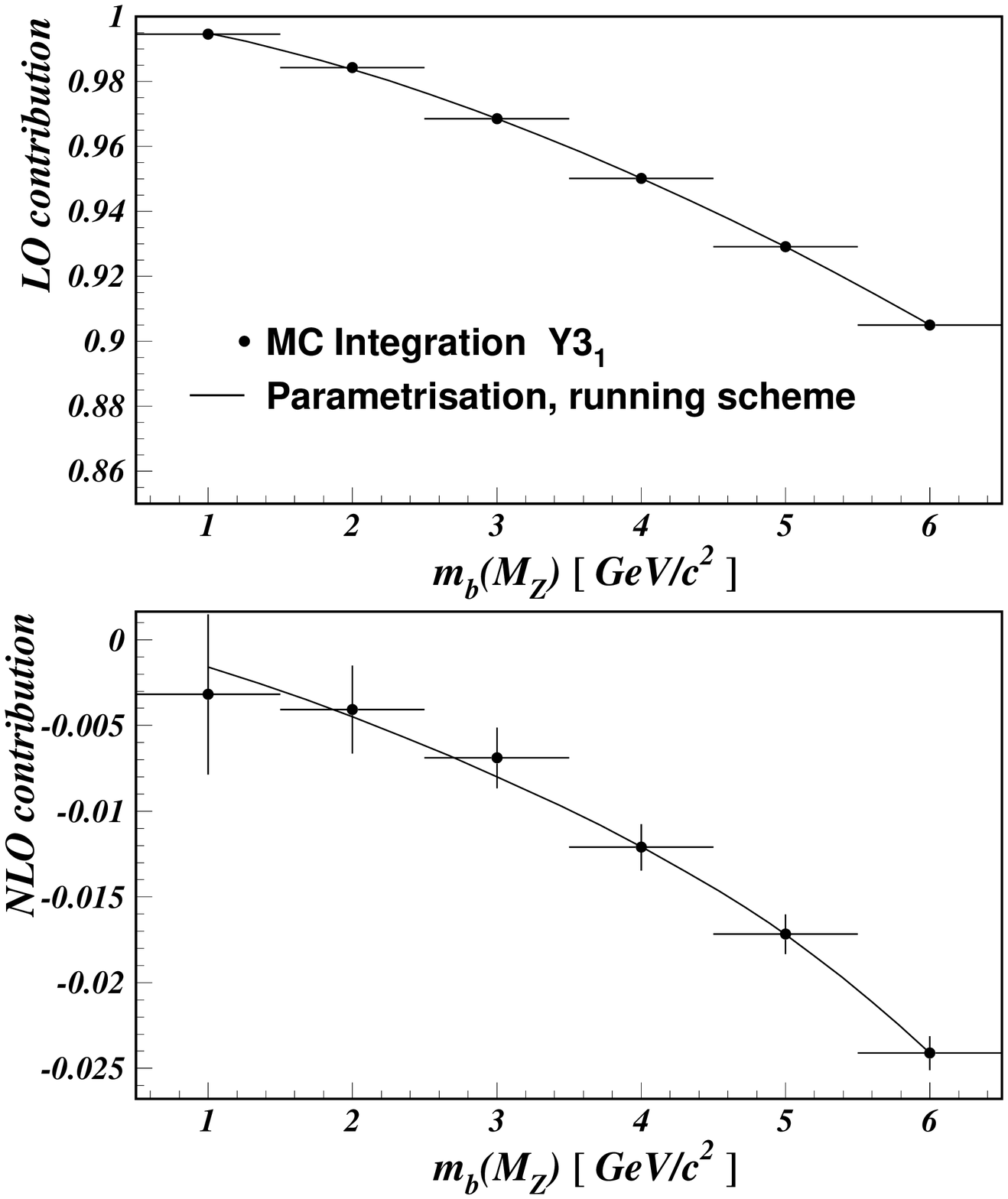}
    \end{tabular}
    \caption{\footnotesize
    Parametrisations of the $b$-quark mass dependence for the 
    LO (top) and NLO (bottom) contributions to $R^{\mathrm{pert}}_{bd}$
    in the pole mass (left) and running mass schemes (right) for
    the first moment of the $y_3$ distribution. The points indicate the result
    of the MC integration, and the full line the parametrisations. 
    The NLO contributions are evaluated using 
    $\alpha_s(M_{\mathrm Z})=0.119$.
    \label{param_thr:fig}}
  \end{center}
\end{figure}

In Fig.~\ref{param_thr:fig} an example of these fits is shown for
the first moment of the $y_3$ distribution. 
With these parametrisations it is possible to estimate the
actual quark mass effects in leading order  and its next-to-leading order
corrections. In Table \ref{lo-nlo:tab} 
a list of these LO and NLO contributions is
given for all variables for both the running and the pole mass schemes.

\begin{figure}[t]
  \begin{center}
    \begin{tabular}{lr}
      \includegraphics[width=8cm]{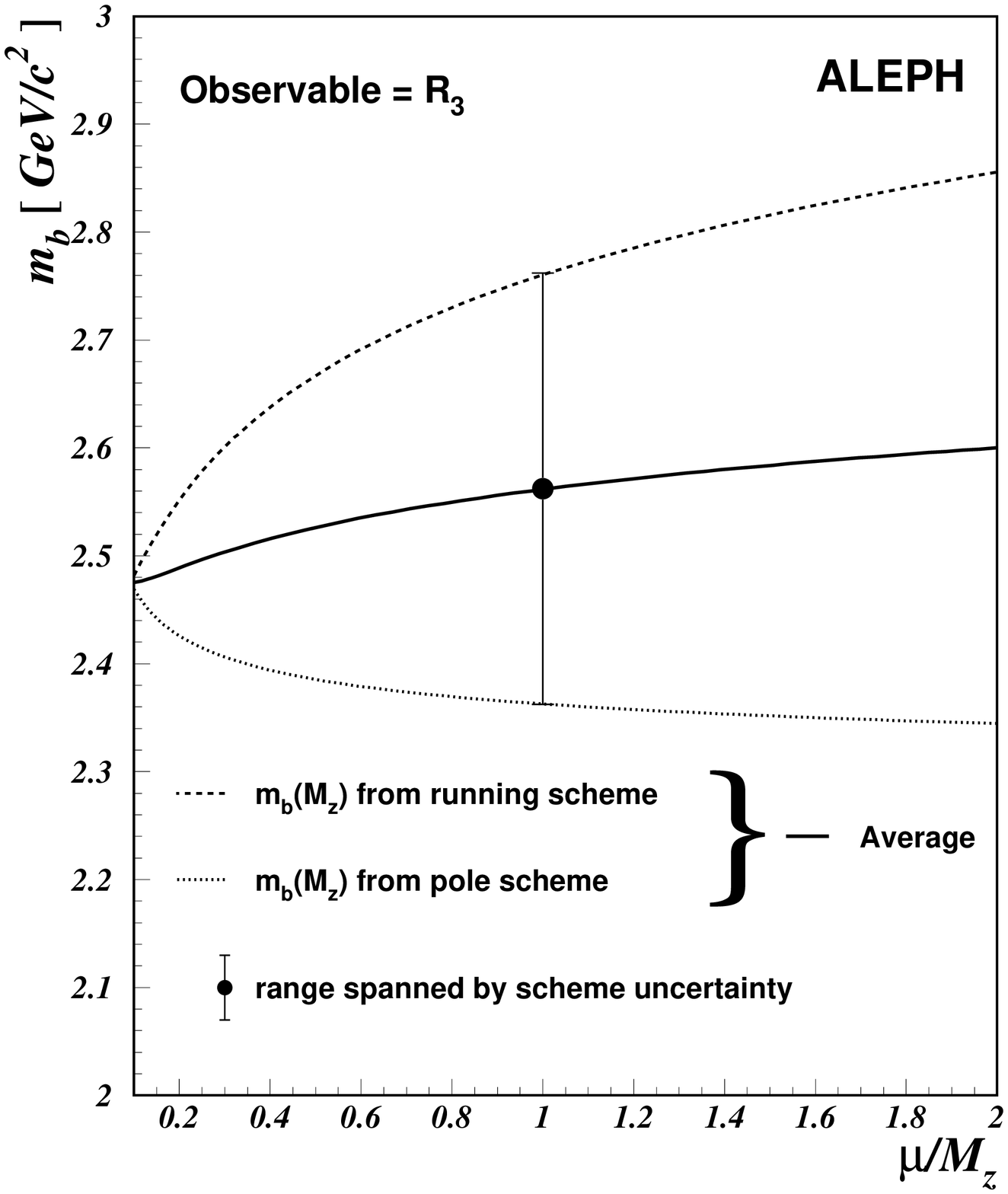} &
      \includegraphics[width=8cm]{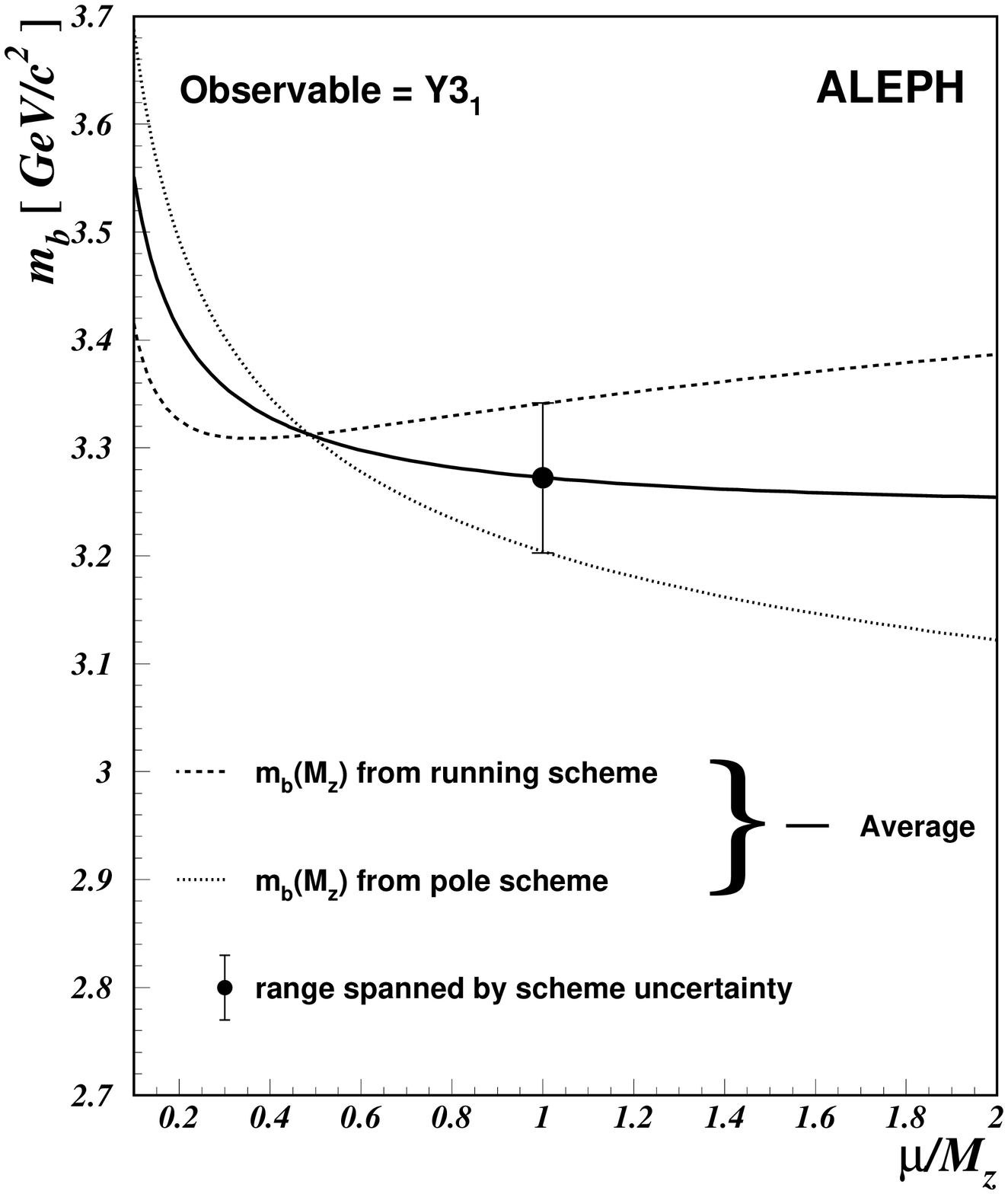} 
    \end{tabular}
    \caption{\footnotesize
    Renormalization scale dependence for the extracted $b$-quark
    mass in the running and pole mass scheme, for the observables
    $R_3$ (left)  and $y_{3_1}$ (right). 
    The error bar spans the range 
    of the scheme uncertainty. The full line indicates the renormalization
    scale dependence of the average computed from the 
    running and pole mass schemes.
    \label{mudep:fig}}
  \end{center}
\end{figure}

Possible evaluations of uncertainties in the mass extraction because
of the limited accuracy of the NLO predictions are discussed next.
Missing higher order corrections are estimated by extracting firstly the
pole mass from the perturbative expression in the pole mass scheme,
then translating that result into a running quark mass $m_b(m_b)$ at the
$b$-mass scale, and finally running this mass up to the $M_{\mathrm Z}$ scale, using
the relation
\be
  \label{running}
   r_b(\mu) = r_b(M_{\mathrm Z}) \left( \frac{\as(M_{\mathrm Z})}{\as(\mu)}\right)^{{-4 \gamma_0}/{\beta_0}} \quad ,
\ee
where 
$\as(\mu) = \as(M_{\mathrm Z})/[1 + \as(M_{\mathrm Z})\beta_0\ln(\mu^2/M_{\mathrm Z}^2)/(4\pi)]$, 
$\beta_0 = 11 - 2n_f/3$,
$n_f = 5$ and $\gamma_0 = 2$.  The world average value
$\asmz=0.119 \pm 0.003$ \cite{Bethke} is used for the strong coupling constant.
At NLO the two methods of mass extraction are equivalent.
However, the sensitivity to higher order contributions is different for the 
two schemes, which can 
then lead to different results. The shift in the mass when using
the pole mass scheme is $-0.14\,\gevcc$ and $-0.4\,\gevcc$ for the
$y_3$ distribution and the three-jet rate, respectively.

In addition, the effects of uncalculated higher order terms can be estimated by
a change in the renormalization scale. For the central value
$\mu = M_{\mathrm Z}$ is employed, and the systematic error is taken to be half of the range
of mass values (average of running and pole mass schemes)
found when varying $\mu$ from 
$0.1 M_{\mathrm Z}$ to $2 M_{\mathrm Z}$. This results in a scale error of $0.15\,\gevcc$
and $0.06\,\gevcc$ for the $y_3$ distribution and the
three-jet rate, respectively.
The behaviour of the extracted mass in the two schemes as a function of the renormalization scale $\mu$ is illustrated in Fig.~\ref{mudep:fig}.

The uncertainty on the world average value 
for the strong coupling constant has a negligible impact,
about $0.01\,\gevcc$,
on the measured $b$-quark mass.


\begin{table}[t]
    \caption{
      \protect\footnotesize 
             Measured $b$-quark mass $m_b(M_{\mathrm Z})$ in the 
             $\overline{\mathrm{MS}}$ scheme with statistical
             (stat), experimental (exp) and hadronization (had)
             uncertainties. 
             In the last column the measured
             pole mass $M_b$ is listed.
            }
  \vspace{0.3cm}
  \begin{center}
    \begin{tabular}{|c|cccc||c|}
      \hline
      \hline
      $O$ & $m_b(M_{\mathrm Z})\, [\gevcc]$ & 
          $\pm$(stat)& 
          $\pm$(exp) & 
          $\pm$(had) & 
          $M_b\, [\gevcc]\;$ \\
      \hline
      \hline
      \rule{0pt}{5.5mm}%
        $R_3$ & 2.76 & 0.28 & 0.28 & 0.62 & 3.65 \\
    $y_{3_1}$ & 3.34 & 0.22 & 0.22 & 0.38 & 4.73 \\[2.5mm]
    \hline
    \hline
    \end{tabular}     
  \end{center}
\label{fullmassres:tab}
\end{table}

\subsection{Results for \boldmath$\mbmz$\unboldmath}
 \label{resultbmass}

Based on the predictions obtained above, the running $b$-quark mass is
determined from the measured ratio $R^{\mathrm{pert}}_{bd}$ for two
observables, the first moment of the $y_3$ distribution and the
three-jet rate.
The results are listed in Table \ref{fullmassres:tab} together with
the statistical and systematic uncertainties, which have been propagated
from the corresponding uncertainties on $R^{\mathrm{pert}}_{bd}$. 
The statistical correlation between the two measurements is 0.66, while
the hadronization uncertainties are almost 100\% correlated.
A similar difference of measured masses was reported in Ref.~\cite{SLACbm1}, 
where different clustering algorithms for the three-jet rate were studied. 
The results obtained for the pole mass by fitting the predictions
in the pole mass scheme are also given in Table \ref{fullmassres:tab}.

As is observed from Table \ref{fullmassres:tab}, 
the first moment of the $y_3$ distribution allows for a mass measurement
with better statistical and experimental accuracy than the three-jet
rate. Furthermore,
the hadronization uncertainty is considerably smaller. 
Therefore the result from this observable 
is quoted as the final $b$-quark mass value.
The theoretical uncertainty is estimated as described in 
Section \ref{theory} by evaluating the impact of the 
uncertainty on the strong coupling constant and the renormalization scale 
variation.
Because the pole and the running mass scheme 
are equivalent at NLO, the average of the values found for the two 
schemes is quoted as the final result, 
and half of the difference is taken as an additional theoretical systematic uncertainty due 
to the scheme ambiguity.
This leads to a measurement of the $b$-quark mass of

\begin{figure}[t]
  \begin{center}
    \includegraphics[width=10cm]{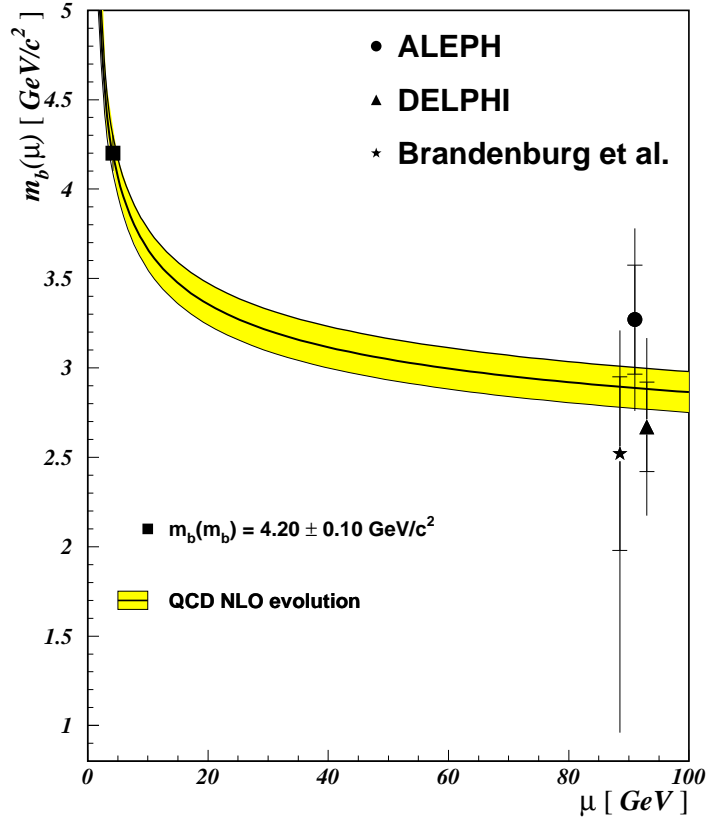}
    \caption{\footnotesize 
     Comparison of the ALEPH result for $m_b(M_{\mathrm Z})$ 
     with the world average value
     of low-energy measurements for $m_b(m_b)$, which is evolved up to
     the $M_{\mathrm Z}$ scale using a two-loop evolution equation with 
     $\alpha_s(M_{\mathrm Z})=0.119\pm0.003$. Also shown are the measurements
     by DELPHI \cite{Delphibm1} and Brandenburg et al.\ \cite{SLACbm1}.
     The inner error bars indicate the quadratic sum of the statistical and
     experimental uncertainties. The three points at the Z pole are
     separated for clarity.
    \label{fig:mb_evol}}
  \end{center}
\end{figure}

\begin{center}
  $m_b(M_{\mathrm Z})  =  \left[ 3.27 \,\pm\, 0.22 (\mathrm{stat}) \,\pm\, 0.22 (\mathrm{exp}) 
           \,\pm\, 0.38 (\mathrm{had})  \,\pm\, 0.16 (\mathrm{theo})\right] \;\gevcc\quad$.  
\end{center}
The result obtained is in agreement with the measurements of 
DELPHI \cite{Delphibm1} and Brandenburg et al.\ \cite{SLACbm1}.
The world average value from low-energy measurements
is $m_b(m_b) = 4.2 \pm 0.1\,\gevcc$ \cite{Pich}.
Using a two-loop running equation for the quark mass,
this average value can be translated to 
$m_b(M_{\mathrm Z}) = 2.89 \pm 0.08\,\gevcc$,
which is also in agreement with the result obtained here.
This comparison is illustrated in Fig.~\ref{fig:mb_evol}, 
together with the other results.
The result for the pole mass is

\begin{center}
$M_b = [ 4.73 \pm 0.29 (\mathrm{stat}) \pm 
                  0.29 (\mathrm{exp})  \pm 0.49 (\mathrm{had}) 
              \pm 0.18 (\mathrm{theo}) ]\,\gevcc$.
\end{center}
In this case the theoretical uncertainty is estimated from the renormalization
scale variation only.

\section{Test of the flavour independence of \boldmath$\alpha_s$\unboldmath}
 \label{flavourindep}

All the previous considerations are based on the assumption of 
flavour independence of the strong coupling constant, which in particular
means that the strong coupling is the same for $b$ quarks as for
light quarks ($l=uds$), i.e., $\alpha_s^b = \alpha_s^{l}$.
However, the measurement of $R^{\mathrm{pert}}_{bd}$ can also be used as a 
test of the flavour independence, if the $b$-quark mass
is known. The ratio of the strong coupling constant
for $b$ and light quark events can be written up to NLO as
\be
 \label{eq:flavindep}
    \frac{\alpha_s^b}{\alpha_s^l} = 
    \left[ R^{\mathrm{pert}}_{bd} - K(r_b(M_{\mathrm Z})) \right] + 
    a_1 \frac{\alpha_s(M_{\mathrm Z})}{\pi} \left[ R^{\mathrm{pert}}_{bd} - 
    K(r_b(M_{\mathrm Z})) - 1 \right]
    \quad , 
\ee
where $K(r_b(M_{\mathrm Z}))$ is the mass correction in Eqn.~\ref{nlogeneral}
and $a_1 = 6.073$ when using the first moment of the $y_3$ distribution
as observable.
Equation \ref{eq:flavindep} is derived assuming $\alpha_s^b = \alpha_s^l (1 + \delta),\;
\delta \ll 1,$ and
neglecting all terms of order $(\alpha_s^l)^2$ and $\alpha_s^l\,\delta$,
so $\alpha_s(M_{\mathrm Z}) = 0.119 \pm 0.003$ can be used
for the coefficient of the next-to-leading order term.
The coefficient $a_1$ was computed with EVENT.
Taking an average $b$-quark mass of $m_b(m_b) = 4.2 \pm 0.1\,\gevcc$
\cite{Pich}, and taking into account the scale
uncertainties and scheme ambiguities as described in Section \ref{theory},
the term $K(r_b(M_{\mathrm Z}))$ amounts to $-0.041 \pm 0.003$. Inserting
the measured value $R^{\mathrm{pert}}_{bd}$ for the observable $y_{3_1}$  
results in

\begin{center}
  $
  \frac{\displaystyle\alpha_s^b}{\displaystyle\alpha_s^l} 
                                = 0.997\,\pm\,0.004\,(\mathrm{stat})
                                       \,\pm\,0.004\,(\mathrm{exp})
                                       \,\pm\,0.007\,(\mathrm{had})
                                       \,\pm\,0.003\,(\mathrm{theo})\nonumber\quad , \nonumber 
  $
\end{center}
which is a confirmation of the flavour independence at a precision
level of 1\%. This constitutes an improvement in precision by a 
factor of two compared to a  previous
ALEPH analysis \cite{ALEPHflavindep}. Recently, similar results were obtained by
other experiments \cite{Delphibm1, OPALflavindep, SLACflavindep}.


\section{Conclusions}
 \label{conclusions}

The effect of the $b$-quark mass has been studied on the 
ratios of observables in $b$ and light quark decays of the Z boson
in $\mathrm{e}^+\mathrm{e}^-$ annihilation. Taking the first moment
of the $y_3$ distribution, which has the smallest hadronization
corrections and systematic uncertainties, a $b$-quark mass of

\begin{center}
$
  m_b(M_{\mathrm Z})  =  \left[ 3.27 \,\pm\, 0.22 (\mathrm{stat}) \,\pm\, 0.22 (\mathrm{exp}) 
           \,\pm\, 0.38 (\mathrm{had})  \,\pm\, 0.16 (\mathrm{theo})\right] \;\gevcc\quad
$
\end{center}

\noindent
is found,
which is in agreement with recent measurements by other experiments 
\cite{Delphibm1, SLACbm1} and with measurements at lower energies.
The three-jet rate defined by the {\tt DURHAM} algorithm with
$y_{\mathrm{cut}}=0.02$ turns out to be more sensitive to the details
of the hadronization modelling. 

The determination
of the ratio of the first moment of the $y_3$ distribution
 in $b$ and light quark events has alternatively
been employed, along with previous measurements of the $b$-quark mass,
to test the flavour independence of the strong coupling
constant with a precision of $1.0\%$.


\section{Acknowledgements}

We would like to thank G.\ Rodrigo and P.\ Nason for providing us with 
theoretical input. Furthermore we would like to thank J.\ Fuster and
S.\ Marti i Garcia for many interesting discussions on the subject.

We wish to thank our colleagues from the accelerator divisions for the successful
operation of LEP. It is also a pleasure to thank the technical personnel of the
collaborating institutions for their support in constructing and maintaining 
the ALEPH experiment. Those of the collaboration not from member states thank CERN
for its hospitality.


\end{document}